\PassOptionsToPackage{hyphens}{url}
\documentclass[aps,pra,superscriptaddress,longbibliography,twocolumn,10pt]{revtex4-2}
\usepackage[utf8]{inputenc}
\usepackage[T1]{fontenc}
\usepackage[english]{babel}
\usepackage{graphicx,amsmath,amssymb,mathtools,dsfont,braket,microtype}
\usepackage[dvipsnames]{xcolor}
\usepackage{hyperref}
\hypersetup{colorlinks=true, linkcolor=MidnightBlue, citecolor=MidnightBlue, urlcolor=MidnightBlue}

\DeclarePairedDelimiterX{\matrixelement}[3]{\langle}{\rangle}{#1 \lvert #2 \rvert #3}
\DeclarePairedDelimiter{\expval}{\langle}{\rangle}
\newcommand{\outerproduct}[2]{\mathinner{\vert #1\rangle\!\langle #2\vert}}

\newcommand{\ee}{\mathrm{e}}
\newcommand{\me}{\mathrm{e}}
\newcommand{\dd}{\mathrm{d}}
\DeclareMathOperator{\Tr}{Tr}
\newcommand{\Nop}{\hat{N}}
\newcommand{\mop}{\hat{m}}

\newcommand{\dv}[2]{\frac{\dd #1}{\dd #2}}
\newcommand{\tuda}{\affiliation{TU Darmstadt, Institute for Applied Physics, Hochschulstra\ss e 4A, D-64289 Darmstadt, Germany}}

\begin{document}
\title{Number fluctuations and particle loss in Dicke superradiance}

\author{Christian Niehof}
\email{cn.niehof@gmail.com}
\tuda
\author{Nico S.~Bassler}
\email{nico.bassler@physik.tu-darmstadt.de}
\tuda
\author{Claudiu Genes}
\email{claudiu.genes@physik.tu-darmstadt.de}
\tuda

\begin{abstract}

Dicke superradiance, the collective spontaneous emission of identical quantum emitters into a common reservoir, produces a burst with a characteristic delay $t_{\mathrm{SR}}$ and a peak intensity that scales quadratically with the emitter number $N$. This strong dependence makes the burst sensitive to how many emitters actually participate. In experiments, however, this number is rarely fixed: it fluctuates between preparations and decreases during emission through atom loss. Here we extend the analytical treatment of superradiance to variable emitter numbers using the Schwinger-boson mapping of the collective spin onto two bosonic modes. In this formulation, collective decay annihilates an atom in the excited-state mode and creates one in the ground-state mode. We find that number fluctuations enhance emission and that the ensemble-averaged emission trace encodes the initial number distribution. We show how to reconstruct second-order correlations from photon counting with finite detection efficiency and temporal resolution, even when the full distribution cannot be resolved. For atom loss at rate $r$, we derive exact results for the population dynamics and show that the peak intensity scaling is governed by the single scaling variable $g_{\mathrm{r}}=r t_{\mathrm{SR}}$. The peak remains quadratic in $N$ for $g_{\mathrm{r}}<1$, follows logarithmically corrected subquadratic power laws for $g_{\mathrm{r}}>1$, and crosses over to linear scaling under strong loss with $g_{\mathrm{r}}\gg1$.
\end{abstract}

\maketitle

\section{Introduction}
\label{sec:introduction}

\begin{figure}[t]
    \centering
    \includegraphics[width=0.82\linewidth]{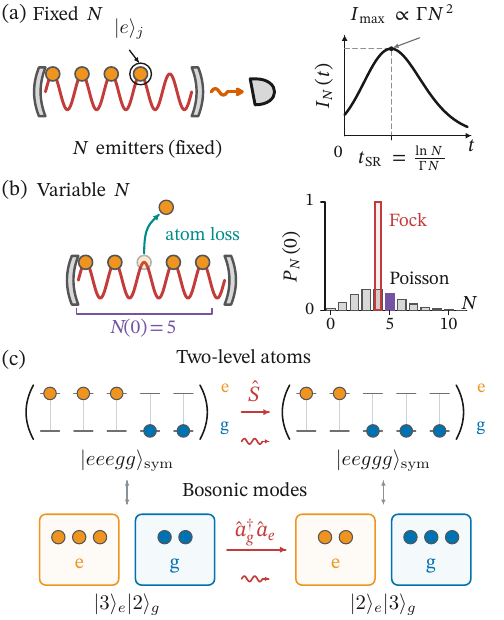}
    \caption{\textbf{Dicke superradiance with number fluctuations and atom loss.}
    (a) Dicke superradiance in a lossy cavity produces enhanced radiation.
    (b) Initial particle-number fluctuations and subsequent atom loss. Poisson and Fock distributions $P_N$, the probabilities of initially preparing $N$ emitters, illustrate the number statistics on the right.
    (c) Exact Schwinger-boson mapping within the fully symmetric Dicke manifold, with $m$ excitations among $N$ atoms, states \(\ket{m}_N=\ket{m}_{\mathrm{e}}\ket{N-m}_{\mathrm{g}}\) and collective lowering operator \(\hat S=\hat a_{\mathrm{g}}^\dagger\hat a_{\mathrm{e}}\).}
    \label{fig:fig1}
\end{figure}

Dicke superradiance describes the collective spontaneous emission of an ensemble of identical two-level atoms into a common electromagnetic reservoir~\cite{dicke1954,gross1982}. In contrast to independent spontaneous emission, where the intensity decays exponentially as $I(t)\propto \exp(-\Gamma t)$, collective emission produces a characteristic superradiant burst with a peak intensity scaling quadratically with the emitter number, \(I_{\mathrm{max}}\propto\Gamma N^2\), reached at a time \(t_{\mathrm{SR}}\sim\ln N/(\Gamma N)\), where $\Gamma$ is the single-emitter decay rate. Such collective emission can be realized, for example, by coupling a collection of atoms to a strongly dissipative cavity [see Fig.~\ref{fig:fig1}(a)], and has been observed in various experimental settings~\cite{norcia20162,liedl2024,ferioli2025,bach2026emergence}.

The strong dependence of collective emission on the emitter number raises the question of how the dynamics of superradiant emission are modified when the particle number is not fixed. In experimental realizations, this may fluctuate due to preparation statistics~\cite{orzel2001,kristensen2019,kruk2025,goban2015}, while atom loss can change the number of participating emitters dynamically [see Fig.~\ref{fig:fig1}(b)]. Here, we employ the Schwinger-boson representation, mapping the fully symmetric collective spin operators onto two bosonic modes corresponding to the excited and ground states, with annihilation operators \(\hat a_{\mathrm{e}}\) and \(\hat a_{\mathrm{g}}\), respectively [see Fig.~\ref{fig:fig1}(c)]. The standard Dicke scenario with fixed particle number corresponds to an initial Fock state $\ket{N}_{\mathrm{e}}$ with $N$ atoms in the excited-state mode, while an initial state with indefinite particle number can be, e.g., a coherent state $\ket{\alpha}_{\mathrm{e}}$~[see Fig.~\ref{fig:fig1}(b)]. Here $\alpha$ is the complex coherent-state amplitude. This leads to a unified description of number-indefinite initial states and processes that couple different particle-number sectors.

Using the Schwinger-boson model, we first investigate how particle-number fluctuations modify superradiant emission, extending previous studies of early-time growth criteria~\cite{robicheaux20212,masson2022}. For initially fully excited ensembles, we use the atom-number operator \(\hat N\) and its normalized second-order correlation \(g_{\mathrm{N}}^{(2)}\) to derive the exact initial-growth criterion \(\langle\hat N\rangle g_{\mathrm{N}}^{(2)}>1\), demonstrating that sufficiently strong number fluctuations can enable initial growth at mean particle numbers below the fixed-\(N\) threshold. We find initial intensity growth for Fock states (\(\langle\hat N\rangle>2\)), coherent states (\(\langle\hat N\rangle>1\)), and thermal states (\(\langle\hat N\rangle>1/2\)), while cat and squeezed-vacuum states exhibit initial growth at any nonzero mean occupation. Moreover, the time-resolved intensity is a weighted sum of the characteristic emission traces from individual particle-number sectors. Collectively scattered light has been used to probe the statistics of quantum gases~\cite{mekhov2007}. Superradiant bursts have also been studied as a means of extracting information about initial spin states~\cite{belliardo2026}. We exploit this dependence to reconstruct the initial particle-number statistics from the ensemble-averaged emission intensity, without requiring higher-order optical correlation measurements. Using simulated photon-counting measurements with finite detection efficiency and temporal resolution, we demonstrate reconstruction of the second-order atom-number correlation \(g_{\mathrm{N}}^{(2)}\), even when the full particle-number distribution is difficult to resolve.

Finally, we investigate the competition between collective emission and continuous atom loss at per-atom rate $r$, complementing previous studies of local decoherence~\cite{shammah2017,bassler2026} and physical atom loss in superradiant lasing~\cite{zhang2018}. We connect the Schwinger-boson description to a microscopic model of local emitter loss and derive exact analytical results for the dynamics within the initial particle-number sector and the evolving particle-number distribution. In the large-\(N\) limit, the competition between superradiant buildup and particle loss produces distinct peak-intensity scaling regimes, governed by the dimensionless ratio \(g_{\mathrm{r}}=r\ln N/(\Gamma N)\). We identify a fully collective regime with quadratic peak intensity, a critical scaling at \(g_{\mathrm{r}}=1\), and a loss-dominated regime exhibiting logarithmically corrected subquadratic power laws, followed by a crossover to linear scaling under sufficiently strong loss.

\section{Short review of Dicke superradiance: Fixed particle number}
\label{sec:fixed_number}
For atoms with excited state \(\ket{e}\) and ground state \(\ket{g}\), the dynamics of a collectively decaying ensemble of quantum emitters are governed by an equation for the density operator $\hat \rho$, known as a master equation
\begin{equation}
    \dot{\hat{\rho}}=\Gamma\Big(\hat{S}\hat{\rho}\hat{S}^{\dagger}-\frac{1}{2}\left\{\hat{S}^{\dagger}\hat{S},\hat{\rho}\right\}_+\Big).
\end{equation}
The bracket denotes anticommutation, and the expression in parentheses defines the Lindblad dissipator $\mathcal D[\hat S]\hat\rho$. Above, $\Gamma$ is the decay rate of an isolated emitter and the collapse operator \(\hat{S}=\sum_{j=1}^{N}\hat{\sigma}_j \) is obtained as a sum over lowering operators \(\hat{\sigma}_j=\ket{g}_j \prescript{}{j}{\bra{e}}\). The full Hilbert space of $N$ two-level systems has dimension $2^N$. The permutation symmetry of the collective collapse operator greatly simplifies the dynamics if the initial state lies in the fully symmetric Dicke manifold: collective emission preserves this subspace, whose dimension is $N+1$. The basis vectors are the Dicke states, denoted by $\ket{m}_N$, where $m=0,\ldots,N$ counts the excitations. The lowering operator acts as
\begin{equation}
    \hat S\ket{m}_N=\sqrt{h_m^{(N)}}\ket{m-1}_N
\end{equation}
where we define the Dicke enhancement factor as
\begin{equation}
h_m^{(N)}=m(N+1-m).
\end{equation}
\begin{figure}[t]
    \centering
    \includegraphics[width=0.9\linewidth]{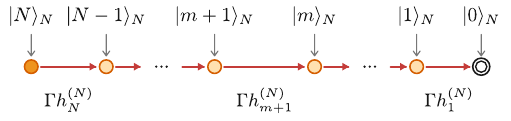}
    \caption{\textbf{Markov chain.} Illustration of Dicke superradiance as a Markov chain of the populations $p_m^{(N)}$ of the Dicke levels $\ket{m}_N$, with transition rates $\Gamma h_m^{(N)}$ from $\ket{m}_N$ to $\ket{m-1}_N$. The state $\ket{N}_N$ is the initial state and $\ket{0}_N$ is the absorbing state.}
    \label{fig:dicke_markov}
\end{figure}
Since the jump operator only transfers population from \(\ket{m}_{N}\) to \(\ket{m-1}_{N}\), the density matrix remains diagonal
\begin{equation}
    \hat{\rho}^{(N)}(t)=\sum_{m=0}^{N}p_m^{(N)}(t)\,\ket{m}_{N}\prescript{}{N}{\bra{m}},
\end{equation}
at all times for any initial state that is diagonal in the Dicke basis. Here we consider the fully excited initial state, $p_m^{(N)}(0)=\delta_{m,N}$, with the Kronecker delta $\delta_{m,N}$. This simplifies the task of solving the master equation: one can directly obtain a set of rate equations for the populations
\begin{equation}
\label{FixedN:recurrence}
\frac{d p_m^{(N)}}{dt}=\Gamma h_{m+1}^{(N)}p_{m+1}^{(N)}
-\Gamma h_m^{(N)}p_m^{(N)},
\end{equation}
describing a Markov chain. This is illustrated in Fig.~\ref{fig:dicke_markov}, where a system starting in the fully excited state, $p_N^{(N)}(0)=1$, eventually reaches the absorbing state at $m=0$.

The dynamics can be directly solved by first integrating the equation for \(m=N\), which gives
\begin{equation}
    p_N^{(N)}(t)
    =\exp[-h^{(N)}_N\Gamma t].
\end{equation}
In subsequent steps, the populations $p_m^{(N)}$ can be obtained by repeated integration of the previous populations $p_{m+1}^{(N)}$, a procedure simplified by the unidirectional Markov-chain structure. Early quantum-statistical and phase-space treatments provided approximate descriptions of superradiant fluctuations and delay times~\cite{bonifacio19711,bonifacio19712,haake1972,degiorgio19711}. Lee subsequently obtained exact time-dependent solutions of the superradiance master equation in 1977~\cite{lee19771,lee19772}. More recent analytical approaches include residue calculus and symbolic quantum trajectories~\cite{holzinger20252,holzinger20251,holzinger2026symbolic}.
For $0\leq m\leq N$, the resulting convolution solution is~\cite{holzinger20251}
\begin{equation}
\begin{aligned}
    p_m^{(N)}(t)={}&\frac{\Gamma^{N-m}N!\,(N-m)!}{m!}\\
    &\times\bigl(e^{-\Gamma h_N^{(N)}t}*\cdots*e^{-\Gamma h_m^{(N)}t}\bigr)(t),
\end{aligned}
\label{eq:fixed_n_convolution}
\end{equation}
where $*$ denotes causal convolution.

For the fixed-$N$ state, the intensity is defined as
\begin{equation}
    I_N(t)=\Gamma \braket{\hat S^\dagger \hat S}=\Gamma\sum_{m=0}^{N}h_m^{(N)}p_m^{(N)}(t).
    \label{eq:fixed_n_intensity}
\end{equation}
The emitted intensity initially scales linearly with the number of emitters, \(I_N(0)=\Gamma N\), whereas the maximum intensity exhibits the characteristic quadratic enhancement \(I_{\mathrm{max}}=I_{\mathrm{SR}}\approx\Gamma N^2/5\) at a time approximately given by $t_{\mathrm{SR}}=\ln N/(\Gamma N)$. At large \(N\), the pulse becomes increasingly narrow, with a characteristic width scaling as \(1/(\Gamma N)\). Since every initially excited atom eventually emits one photon, the integrated intensity satisfies \(\int_0^\infty I_N(t)\,dt=N\).

\section{Schwinger boson mapping: Variable particle number}
\label{sec:variable_number}

\begin{figure}
    \centering
    \includegraphics[width=\linewidth]{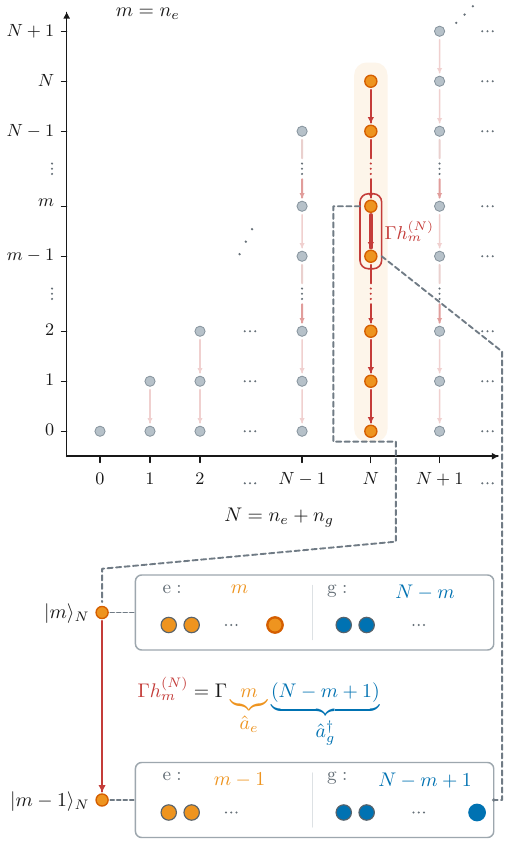}
    \caption{\textbf{Disconnected Dicke triangle.} We illustrate variable particle number on a Dicke triangle where horizontal and vertical axes represent \(N=n_{\mathrm{e}}+n_{\mathrm{g}}\) and \(m=n_{\mathrm{e}}\), where $n_{\mathrm{e}}$ and $n_{\mathrm{g}}$ count excited- and ground-state atoms. Each vertical chain is a fixed-\(N\) sector, which evolves independently under collective emission. The highlighted chain and magnified transition illustrate \(|m\rangle_N\rightarrow|m-1\rangle_N\), corresponding to decoupled population dynamics. The rate \(\Gamma h_m^{(N)}\) follows from the squared bosonic matrix element, with factors associated with annihilation in the excited mode and creation in the ground mode.}
    \label{fig:triangle_decoupled}
\end{figure}

In experimental realizations, however, the number of emitters participating in collective emission is generally not perfectly fixed. Fluctuations in the initial atom number can arise from imperfect preparation or intrinsic number statistics~\cite{orzel2001,kristensen2019,kruk2025}. We thus consider general initial states of fully excited atoms, potentially mixed, with contributions from different particle-number sectors,
\begin{equation}
    \hat{\rho}(0)=\sum_{N,N'=0}^{\infty}\hat{\rho}^{(N,N')}(0),
\end{equation}
where
\begin{equation}
    \hat{\rho}^{(N,N')}(0)=\sum_{m=0}^{N}\sum_{m'=0}^{N'}\rho_{m,m'}^{(N,N')}(0)\ket{m}_{N}\prescript{}{N'}{\bra{m'}}.
\end{equation}

The matrix elements $\rho_{m,m'}^{(N,N')}(0)$ might be from a coherent state, a thermal mixture, an even cat state (hereafter simply a cat state) or any other distribution.

To treat variable particle number $N$ we adopt the Schwinger-boson representation of the collective spin operators~\cite{schwinger1952angular,silva2022permutational}, introducing two bosonic modes, \(\hat{a}_{\mathrm{e}}\) and \(\hat{a}_{\mathrm{g}}\), which represent atoms in the excited and ground states, respectively. The symmetric Dicke states are identified with two-mode Fock states according to
\begin{equation}
    \ket{m}_N=\ket{n_{\mathrm{e}}=m,n_{\mathrm{g}}=N-m}.
    \label{eq:schwinger_basis}
\end{equation}
Their action on the corresponding number states is the usual bosonic one,
\begin{equation}
    \hat{a}_{\mathrm{e}}\ket{n_{\mathrm{e}},n_{\mathrm{g}}}=\sqrt{n_{\mathrm{e}}}\ket{n_{\mathrm{e}}-1,n_{\mathrm{g}}},
\end{equation}
and analogously for the ground-state mode. The collective spin operators are then mapped onto bilinear combinations of the two bosonic operators,
\begin{equation}
    \hat{S}=\hat{a}_{\mathrm{g}}^{\dagger}\hat{a}_{\mathrm{e}},
    \qquad
    \hat{S}^{\dagger}=\hat{a}_{\mathrm{e}}^{\dagger}\hat{a}_{\mathrm{g}},
    \qquad
    \hat{S}_{\mathrm{z}}=\frac{1}{2}\left(
    \hat{a}_{\mathrm{e}}^{\dagger}\hat{a}_{\mathrm{e}}
    -
    \hat{a}_{\mathrm{g}}^{\dagger}\hat{a}_{\mathrm{g}}
    \right).
\end{equation}

In this representation, the action of the collective lowering operator \(\hat{S}\) has a particularly simple interpretation: one excited-state atom is annihilated and one ground-state atom is created. In particular,
\begin{equation}
    \hat{S}\ket{m}_N=\sqrt{h_m^{(N)}}\ket{m-1}_N,
\end{equation}
recovering the Dicke enhancement factors \(h_m^{(N)}\) as the squared matrix elements of the two bosonic operators.

The total atom-number and excitation-number operators are
\begin{equation}
\begin{aligned}
    \Nop&=\hat{a}_{\mathrm{e}}^{\dagger}\hat{a}_{\mathrm{e}}+\hat{a}_{\mathrm{g}}^{\dagger}\hat{a}_{\mathrm{g}},\\
    \mop&=\hat{a}_{\mathrm{e}}^{\dagger}\hat{a}_{\mathrm{e}}.
\end{aligned}
\label{eq:number_operators}
\end{equation}
Their eigenvalues are denoted by \(N\) and \(m\), respectively. The total atom-number operator satisfies
\begin{equation}
    [\Nop,\hat{S}]=[\Nop,\hat{S}^{\dagger}]=0.
\end{equation}
Consequently, Dicke superradiance with a fixed particle number \(N\) is recovered by restricting the two-mode Fock space to the subspace \(n_{\mathrm{e}}+n_{\mathrm{g}}=N\). As illustrated in Fig.~\ref{fig:triangle_decoupled}, the usual Dicke ladder is then represented by a single fixed-\(N\) chain of the two-mode Fock lattice. Starting from the fully excited state, the superradiant cascade corresponds to the successive conversion in \((n_{\mathrm{e}},n_{\mathrm{g}})\) coordinates,
\begin{equation}
    (N,0)\rightarrow(N-1,1)\rightarrow\cdots\rightarrow(0,N),
\end{equation}
while the total particle number remains unchanged.

The main advantage of the Schwinger-boson representation becomes apparent once the particle number is not fixed. Superpositions or mixtures with different total atom numbers, such as an initial coherent state of the excited-state mode \(\ket{\alpha_{\mathrm{e}}}\), occupy several fixed-\(N\) sectors of the same two-mode Fock space, as illustrated in Fig.~\ref{fig:triangle_decoupled}. Each fixed-\(N\) component undergoes its own Dicke cascade, while the complete dynamics is described within a single common Hilbert space. This formulation is therefore particularly natural for number-indefinite initial states and, more generally, for processes such as particle loss that couple different particle-number sectors.

To compute the emitted intensity, the task can be further simplified by noticing that for observables that commute with the total atom number, \([\hat{O},\Nop]=0\), only the diagonal \(N=N'\) blocks contribute to expectation values. Defining the particle-number probabilities \(P_N=\Tr\hat{\rho}^{(N,N)}(0)\) and the normalized fixed-\(N\) density matrices \(\hat{\rho}^{(N)}(t)=\hat{\rho}^{(N,N)}(t)/P_N\) for \(P_N>0\), we obtain
\begin{equation}
    \langle\hat{O}(t)\rangle=\sum_{N=0}^{\infty}P_N\Tr\{\hat{O}\hat{\rho}^{(N)}(t)\}.
\end{equation}

\begin{figure}[ht]
    \centering
    \includegraphics[width=0.92\columnwidth]{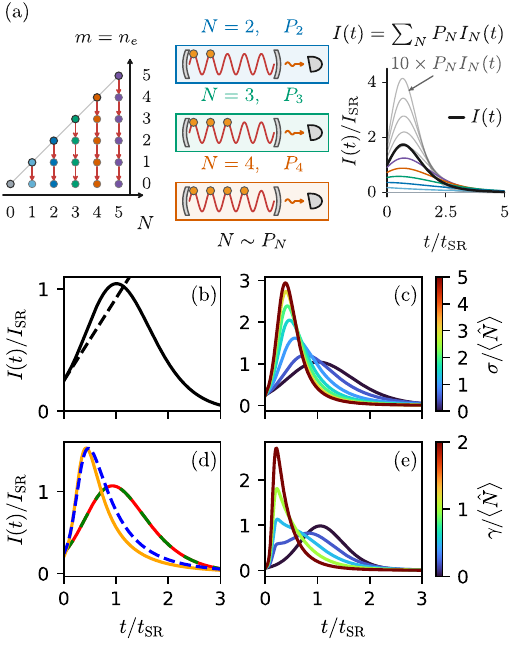}
    \caption{\textbf{Variable number intensity traces.}
        Intensity profiles for different initial distributions of ensembles of fully excited atoms.
        The reference scales $I_{\mathrm{SR}}$ and $t_{\mathrm{SR}}$ are evaluated at $N=\braket{\Nop}$.
        (a) illustrates number-averaged emission from independent fixed-$N$ sectors.
        The cavities on the right represent ensembles with fixed emitter number $N$, each evolving along a separate Markov chain in the triangle on the left.
        The gray and colored curves show the weighted contributions $P_N I_N(t)$ magnified by a factor of 10 for visibility.
        The black curve shows the actual total $I(t)=\sum_N P_N I_N(t)$, obtained by summing the unmagnified contributions.
        (b) shows the profile of a Fock state for $N=20$ as a solid line and its initial slope as a dashed line.
        (c) displays different truncated Gaussian distributions using a cutoff of $N_{\mathrm{max}}=100$ for a mean particle number of $\braket{\Nop}=20$ with colors indicating their respective standard deviations $\sigma$ before truncation, according to the colorbar.
        (d) compares the coherent state (red), the cat state (green), the thermal state (blue), and the squeezed state (orange) for a mean particle number of $\braket{\Nop}=20$.
        (e) exhibits the features of the truncated Lorentzian distribution with half-width $\gamma$ for a mean particle number of $\braket{\Nop}=100$ using $N_{\mathrm{max}}=1000$.
        The colors of the intensity represent the relative scale $\gamma/\braket{\Nop}$.}
    \label{fig3}
\end{figure}

In particular, the emitted intensity is
\begin{equation}
    I(t)=\sum_{N=0}^{\infty}P_NI_N(t),
    \label{eq:number_averaged_intensity}
\end{equation}
where \(I_N(t)=\Gamma\Tr\{\hat{S}^{\dagger}\hat{S}\hat{\rho}^{(N)}(t)\}\). Thus, the intensity is simply the average of the intensity traces from the fixed-$N$ sectors weighted by $P_N$ as illustrated in Fig.~\ref{fig3}(a). The $N$-emitter Dicke cascades are illustrated by the Dicke triangle on the left and the corresponding independent cavity systems on its right:
\begin{equation}
    I(t)=\Gamma\sum_{N=0}^{\infty}P_N\sum_{m=0}^{N}h_m^{(N)}p_m^{(N)}(t).
\end{equation}

The conservation of total particle number implies that coherences between different particle-number sectors neither contribute to number-conserving observables nor influence the dynamics of the diagonal \(N=N'\) blocks. They can therefore be discarded entirely for the observables considered here. Thus the dynamics reduce exactly to a statistical mixture of independent fixed-\(N\) Dicke cascades, with the emitted intensity given by Eq.~\eqref{eq:number_averaged_intensity}. For initially fully excited ensembles, the dynamics within each sector is uniquely determined by \(N\), such that the initial particle-number distribution \(P_N\) completely specifies the resulting intensity.

\section{Initial number statistics and superradiant emission}\label{sec:initial_statistics}

We consider Fock, coherent, cat, thermal, and squeezed-vacuum number statistics, whose distributions and corresponding moments are summarized in Table~\ref{tab:number_distributions}.

\begin{table*}[!t]
    \centering
    \renewcommand{\arraystretch}{1.35}
    \setlength{\tabcolsep}{1.5pt}
    \begin{tabular}{l|l|c|c|c|c}
        \hline\hline
        Distribution & Number distribution $P_N$ & $\langle \Nop^2\rangle$ & $\langle \Nop\rangle$ & Growth condition & Condition on $\langle \Nop\rangle$ \\
        \hline
        Fock
        & $\delta_{N,N_0}$
        & $N_0^2$
        & $N_0$
        & $N_0>2$
        & $\langle \Nop\rangle>2$
        \\
        Coherent
        & $\me^{-|\alpha|^2}\dfrac{|\alpha|^{2N}}{N!}$
        & $|\alpha|^4+|\alpha|^2$
        & $|\alpha|^2$
        & $|\alpha|>1$
        & $\langle \Nop\rangle>1$
        \\
        cat
        & $P_{2k}=\operatorname{sech}(|\alpha|^2)\dfrac{|\alpha|^{4k}}{(2k)!},\quad P_{2k+1}=0$
        & $|\alpha|^4+|\alpha|^2\tanh|\alpha|^2$
        & $|\alpha|^2\tanh|\alpha|^2$
        & $|\alpha|>0$
        & $\langle \Nop\rangle>0$
        \\
        Thermal
        & $(1-\me^{-\delta})\me^{-\delta N}$
        & $\dfrac{\me^\delta+1}{(\me^\delta-1)^2}$
        & $\dfrac{1}{\me^\delta-1}$
        & $\delta<\ln 3$
        & $\langle \Nop\rangle>\dfrac{1}{2}$
        \\
        Squeezed vacuum
        & $P_{2k}=\dfrac{\tanh^{2k}s}{\cosh s}\dfrac{(2k)!}{4^k(k!)^2},\quad P_{2k+1}=0$
        & $3\sinh^4 s+2\sinh^2 s$
        & $\sinh^2 s$
        & $s>0$
        & $\langle \Nop\rangle>0$
        \\
        \hline\hline
    \end{tabular}
    \caption{Particle-number distributions, their first two moments, and the condition for an initially increasing Dicke intensity, $\dot I(0)=\Gamma^2[\langle\Nop^2\rangle-2\langle\Nop\rangle]>0$. Here $N_0$ is the Fock occupation, $\alpha$ the complex coherent-state amplitude, $s\geq0$ the real squeezing parameter, and $k=0,1,\ldots$. The dimensionless inverse temperature is $\delta=\hbar\omega_0/(k_{\mathrm B}T)>0$, with transition angular frequency $\omega_0$ and temperature $T$.}
    \label{tab:number_distributions}
\end{table*}

An important question in Dicke superradiance is under which conditions an ensemble exhibits an initially increasing emission intensity. Early-time growth criteria have been investigated for atomic arrays~\cite{robicheaux20212} and connected to the second-order photon correlation function \(g_{\mathrm{ph}}^{(2)}(0)\)~\cite{masson2022}. Related studies have considered degenerate quantum matter~\cite{lyne2026} and competing decoherence processes~\cite{bassler2026}. Here, we extend the initial-growth criterion to initially fully excited ensembles with fluctuating particle numbers, relating the short-time behavior of the emitted intensity directly to moments of the initial number distribution. We find in App.~\ref{app:moments}
\begin{equation}
    \frac{\dot I(0)}{\Gamma^2}
    =\braket{\Nop^2}-2\braket{\Nop}
\end{equation}
Thus, particle-number fluctuations directly enhance the initial growth rate. At fixed integer mean particle number, the Fock state minimizes $\dot I(0)$, while broader number distributions can exhibit an initially increasing intensity at substantially smaller mean occupation. The corresponding conditions for the distributions considered here are summarized in Table~\ref{tab:number_distributions}. Higher-order derivatives similarly probe higher moments of the number distribution; we return to this observation in Sec.~\ref{sec:reconstruction}, with explicit expressions given in App.~\ref{app:moments}. The initial-growth condition can equivalently be expressed using the normalized second-order correlation function of the counting statistics $g_{\mathrm{N}}^{(2)}=
{\expval{\Nop(\Nop-1)}}/{\expval{\Nop}^2}$, yielding
\begin{equation}
    \expval{\Nop}g_{\mathrm{N}}^{(2)}>1.
\end{equation}
The atom-number correlation function is connected to the normalized second-order photonic correlation function via \(g_{\mathrm{ph}}^{(2)}(0)={\langle\hat S^{\dagger 2}\hat S^2\rangle}/{\langle\hat S^\dagger\hat S\rangle^2}=2g_{\mathrm{N}}^{(2)}\), also yielding the condition $g_{\mathrm{ph}}^{(2)}(0)>2/\expval{\Nop}$. For fixed particle number, $g_{\mathrm{ph}}^{(2)}(0)=2(1-1/N)$, so both this condition and the photon-correlation criterion \(g_{\mathrm{ph}}^{(2)}(0)>1\) discussed in Ref.~\cite{masson2022} reduce to $N>2$. For fluctuating particle numbers, however, photon bunching alone is not sufficient to ensure an initially increasing intensity, and the criterion additionally depends on the mean emitter number.

The thresholds show a clear distribution dependence. A Fock state requires $\braket{\Nop}>2$ for the intensity to initially increase, whereas the threshold is reduced to $\braket{\Nop}>1$ for a coherent state and $\braket{\Nop}>1/2$ for a thermal state. For the cat and squeezed-vacuum states, $\dot I(0)>0$ for any nonzero mean occupation. Thus, an initially increasing superradiant intensity is not determined by the mean particle number alone, but can be enabled by sufficiently strong particle-number fluctuations.

The cat and squeezed-vacuum distributions contain only even particle numbers, while the thermal distribution exhibits substantially larger number fluctuations than a Poissonian distribution at the same mean occupation [see App.~\ref{app:dist}]. For comparison, we also introduce phenomenological Gaussian and Lorentzian number distributions with center $N_0$, Gaussian standard deviation $\sigma>0$, and Lorentzian half-width $\gamma>0$,
\begin{equation}
    P_N\propto\exp\left[-\frac{(N-N_0)^2}{2\sigma^2}\right],
    \qquad
    P_N\propto\frac{1}{(N-N_0)^2+\gamma^2},
\end{equation}
respectively. In the numerical calculations these distributions are truncated at a finite particle number $N_{\mathrm{max}}>2N_0$. The truncated distributions are weighted against the vacuum such that their mean coincides with the desired parameter $N_0$ and the density matrix has trace one. As a consequence, the variance and higher moments are different from the original distributions, but their mean is fixed at $N_0$. The Lorentzian case is particularly sensitive to the cutoff, since both its first and second moments diverge without truncation. For sufficiently large cutoffs, the vacuum-adjusted distributions therefore exhibit an initially increasing intensity at arbitrarily small positive mean occupation.

Fig.~\ref{fig3} shows how these short-time differences extend to the complete intensity dynamics. Fig.~\ref{fig3}(b) illustrates the fixed-$N$ Fock state case, while Fig.~\ref{fig3}(d) compares several initial states with the same mean particle number. The coherent and cat states yield similar intensity profiles at large occupation. Although their number distributions retain different parity structures, the intensity becomes increasingly insensitive to these fine features, consistent with the growing overlap of neighboring fixed-\(N\) intensity traces. By contrast, the broader thermal and squeezed-vacuum distributions produce an earlier and higher intensity maximum. This reflects the nonlinear weighting of different particle-number sectors: sectors with larger $N$ radiate more strongly, with a peak intensity scaling as $N^2$, and peak earlier, at $t_N\sim \ln N/(\Gamma N)$.

Panels (c) and (e) of Fig.~\ref{fig3} isolate this effect by continuously broadening phenomenological particle-number distributions at fixed mean occupation. Increasing the width of the Gaussian distribution progressively enhances the contribution of larger-$N$ sectors, shifting the intensity maximum to earlier times and increasing its amplitude, while smaller-$N$ sectors contribute a longer-time tail. The effect is even more pronounced for the heavy-tailed Lorentzian distribution. Since both its first and second moments diverge without truncation, the comparison is restricted to finite-cutoff distributions with adjusted mean occupation. In particular, the short-time behavior remains sensitive to the imposed upper particle-number cutoff. These differences in the temporal emission profiles suggest that the intensity itself can serve as a probe of the initial atom-number distribution, as we investigate in the following section.

\section{Reconstruction of number statistics from the burst}
\label{sec:reconstruction}

Collective emission encodes initial atom-number fluctuations in the time dependence of the ensemble-averaged intensity. The reconstruction of number statistics from indirect measurements has been investigated in several settings~\cite{olivares2007,zambra2006,weyland2025}, including the extraction of atomic statistics from collectively scattered light~\cite{mekhov2007}. Related approaches have also explored the extraction of information about initial spin states from superradiant bursts~\cite{belliardo2026}. More generally, the time evolution of a single expectation value can provide sufficient information for quantum-state tomography under suitable dynamics~\cite{silberfarb2005,rall2026,peruzzo2026}.

\begin{figure}[!t]
    \centering
    \includegraphics[width=\linewidth]{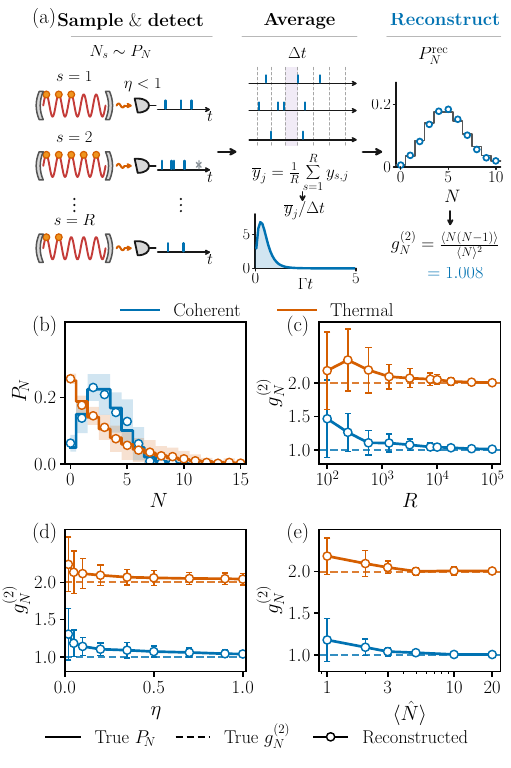}
\caption{\textbf{Reconstruction of particle-number statistics from Dicke superradiance.}
(a) Measurement protocol: each of $R$ preparations samples $N_s\sim P_N$ and excites all $N_s$ emitters. Photon records, collected at known efficiency $\eta$ and filtered by the known response, are binned at width $\Delta t$ and averaged. Fitting fixed-$N$ kernels to the mean record yields reconstructed probabilities $P_N^{\mathrm{rec}}$ and $g_{\mathrm{N}}^{(2)}$.
(b) True probabilities (steps) and mean reconstructions (open circles) for coherent (blue) and thermal (orange) states; only $N\leq15$ is displayed.
(c--e) Mean reconstructed $g_{\mathrm{N}}^{(2)}$ versus preparation number $R$, detection efficiency $\eta$, and mean occupation $\expval{\Nop}$, respectively; dashed lines indicate the true values. Error bars and shading show one sample standard deviation across 30 reconstructions using independent photon records. Preparations in (c) are nested within runs.
The mean occupation is $\expval{\Nop}=3$ unless varied, and panels (b), (d), and (e) use $R=10^4$.
The response time $\tau$ and acquisition duration $T_{\mathrm{acq}}$ satisfy $\Gamma\tau=0.05$ and $\Gamma T_{\mathrm{acq}}=5$; $\Gamma\Delta t=0.01$ and $\eta=0.9$ unless varied. Reconstruction uses nonnegative, normalized least squares with a second-difference penalty $\lambda=10^{-6}$.
Both distributions share an adaptive cutoff $N_{\mathrm{max}}\geq150$ satisfying $\sum_{N>N_{\mathrm{max}}}P_N<10^{-10}$; the thermal tail sets $N_{\mathrm{max}}=241$ and $471$ at mean occupations $10$ and $20$, respectively.}
\label{fig:reconstruction}
\end{figure}

Here, we exploit the known fixed-\(N\) Dicke intensity curves to reconstruct initial atom-number statistics from the ensemble-averaged emission trace. We focus on the second-order atom-number correlation function \(g_{\mathrm{N}}^{(2)}\), which can be reconstructed even when the full number distribution is difficult to resolve. We illustrate the approach for coherent and thermal initial number distributions using finite-statistics intensity records, including finite detection efficiency $\eta$ and a known detector response. This is illustrated schematically in Fig.~\ref{fig:reconstruction}(a).

With complete pulse collection and no emitter loss, shot-to-shot integrated photon counts also give \(g_{\mathrm{N}}^{(2)}\): independent photon loss multiplies the \(k\)-th factorial moment by \(\eta^k\), leaving normalized factorial correlations unchanged. Such counting requires neither efficiency calibration nor knowledge of the temporal detector response. However, resolving the total photon number in each dense burst can be impractical, and many experiments provide only the ensemble-averaged time profile~\cite{liedl2024,mlynek2014}. Our method extracts number correlations from this trace without shot-to-shot fluctuation information, at the cost of requiring a known detection efficiency and detector response.

\paragraph*{Derivative-based reconstruction.} The dependence of the emitted intensity on the initial particle-number distribution also provides a direct way of extracting its moments from the short-time dynamics. For an initially fully excited ensemble, successive time derivatives of the intensity at \(t=0\) are polynomials in the moments of the initial number distribution. The first few derivatives read
\begin{equation}
\frac{I(0)}{\Gamma}=\langle \Nop\rangle,\qquad
\frac{\dot I(0)}{\Gamma^2}
=\langle \Nop^2\rangle-2\langle \Nop\rangle,
\end{equation}
and
\begin{equation}
\frac{\ddot I(0)}{\Gamma^3}
=\langle \Nop^3\rangle-8\langle \Nop^2\rangle+8\langle \Nop\rangle.
\end{equation}

More generally, the normalized derivative $I^{(k)}(0)/\Gamma^{k+1}$ contains moments up to order \(k+1\), with the highest moment \(\langle \Nop^{k+1}\rangle\) appearing linearly with unit coefficient. The resulting hierarchy is therefore triangular: \(I(0)\) determines the mean particle number, \(\dot I(0)\) then determines the second moment, and each subsequent derivative determines one additional moment once the lower moments are known. In principle, knowledge of the complete short-time Taylor expansion of the radiation intensity therefore determines the full sequence of moments of the initial particle-number distribution. Explicit expressions for the first several orders and the corresponding inverse relations are given in App.~\ref{app:moments}.

\paragraph*{Reconstruction based on the intensity curve.} Using Eq.~\eqref{eq:number_averaged_intensity}, we truncate the initial number distribution at $M=N_{\mathrm{max}}$ and sample the intensity at $M$ times $t_1,\ldots,t_M$. The probabilities $P_1,\ldots,P_M$ then obey the linear system
\begin{equation}
\begin{pmatrix}
I(t_1)\\
I(t_2)\\
\vdots\\
I(t_M)
\end{pmatrix}
=
\begin{pmatrix}
I_1(t_1) & I_2(t_1) & \cdots & I_M(t_1)\\
I_1(t_2) & I_2(t_2) & \cdots & I_M(t_2)\\
\vdots & \vdots & \ddots & \vdots\\
I_1(t_M) & I_2(t_M) & \cdots & I_M(t_M)
\end{pmatrix}
\begin{pmatrix}
P_1\\
P_2\\
\vdots\\
P_M
\end{pmatrix}.
\end{equation}
Since $I_0(t)=0$, the vacuum probability $P_0$ follows from normalization. As \(N\) increases, neighboring fixed-\(N\) intensity curves become increasingly similar, making reconstruction of the full particle-number distribution progressively ill-conditioned. This can be quantified through their overlap, which approaches unity in the large-\(N\) limit, as shown in App.~\ref{app:intensity_cond}.

\paragraph*{Demonstration of reconstruction under realistic conditions.}
The preceding reconstruction methods rely on accurately known intensities or derivatives thereof. With finite counting statistics (shot noise), numerical differentiation amplifies fluctuations, while direct matrix inversion is sensitive to the increasingly similar neighboring intensity kernels discussed above. As a more stable alternative, we use a least-squares optimization with positivity and normalization constraints on the initial number probabilities. Additionally, we include a weak smoothness penalty to suppress oscillatory reconstructed distributions that produce nearly identical intensity traces~\cite{hansen2010}.

For recording bins $j$ starting at $t_j$, with width $\Delta t$ within a fixed acquisition interval $[0,T_{\mathrm{acq}}]$, we define the bin-integrated intensity kernels
\begin{equation}
K_{jN}=\int_{t_j}^{t_j+\Delta t}\dd t\int_0^t\dd s\,f_\tau(t-s)I_N(s),
\end{equation}
where $f_\tau(t)=\Theta(t)e^{-t/\tau}/\tau$ describes a causal detector response with response time $\tau$ and the Heaviside step function $\Theta(t)$; $f_0(t)=\delta(t)$, with the Dirac delta, corresponds to an instantaneous response. With a numerical particle-number cutoff $N_{\mathrm{max}}$, the observed bin-integrated signal per preparation, $y_j$, has expectation $\mathbb{E}[y_j]=\eta\sum_{N=0}^{N_{\mathrm{max}}}K_{jN}P_N$, where $\eta$ is the detection efficiency. Both the detection efficiency and the detector response are assumed to be known. Without independent calibration of the detection efficiency, the mean intensity trace does not in general uniquely determine the initial number distribution, since changes in efficiency can be compensated by changes in the probabilities.

To generate synthetic observations, we sample $N$ from the prescribed initial distribution and generate emission times from the Markov chain of Eq.~\eqref{FixedN:recurrence}, with transition rates $\Gamma h_m^{(N)}$. Each emitted photon is independently registered with probability $\eta$, and each registered photon contributes a time-shifted response $f_\tau$. The resulting signals are integrated over the recording bins and averaged over $R$ independent preparations. This includes fluctuations in the initial particle number, emission times, and detection events, whose standard deviation in the averaged signal decreases as $R^{-1/2}$.

Collecting $P_N$ and $y_j$ into vectors $\mathbf P$ and $\mathbf y$, and $K_{jN}$ into the matrix $K$, we obtain the reconstructed distribution from the convex quadratic problem
\begin{equation}
\mathbf P^{\mathrm{rec}}
=
\underset{\substack{P_N\geq0\\
\sum_{N=0}^{N_{\mathrm{max}}}P_N=1}}
{\operatorname{argmin}}
\left[
\|\eta K\mathbf P-\mathbf y\|_2^2
+
\lambda_{\Delta t}\|D_2\mathbf P\|_2^2
\right],
\end{equation}
where $\|\cdot\|_2$ is the Euclidean norm, $\lambda_{\Delta t}\geq0$ is the smoothing strength, and $(D_2\mathbf P)_N=P_{N+1}-2P_N+P_{N-1}$ for $1\leq N<N_{\mathrm{max}}$. The smoothing term favors slowly varying probabilities without imposing a particular distribution family. When varying the bin width, we use $\lambda_{\Delta t}=\lambda_{\mathrm{ref}}\Delta t/\Delta t_{\mathrm{ref}}$, where $\lambda_{\mathrm{ref}}$ is the strength at reference bin width $\Delta t_{\mathrm{ref}}$, to account for the leading bin-width dependence of the squared integrated residuals for a temporally resolved trace. In numerical checks for the smooth distributions considered here, reconstructions with $\lambda_{\Delta t}=0$ and with the finite smoothing strength used here yield essentially the same low-order moments, although the reconstructed distributions can differ substantially, see App.~\ref{app:rec}.

We use the normalized second-order correlation function of the initial number statistics, $g_{\mathrm{N}}^{(2)}$, which equals $g_{\mathrm{N}}^{(2)}=1$ for a Poisson distribution and $g_{\mathrm{N}}^{(2)}=2$ for a thermal distribution, independently of the nonzero mean occupation. Although the preceding noiseless reconstruction methods provide formal access to the initial number distribution, its detailed reconstruction becomes increasingly difficult as the mean particle number grows, as indicated by Eq.~\eqref{eq:overlap}. In particular, the increasing overlap between neighboring intensity kernels limits the resolution of fine features in the number distribution. An illustrative example is the parity structure of a cat state, for which only even particle numbers are occupied, as shown in App.~\ref{app:rec}. Resolving this alternating occupation pattern requires information beyond a few low-order moments and is particularly sensitive to statistical fluctuations; the smoothness penalty also disfavors such structure. Consequently, accurate reconstruction of the full distribution, or even its parity, can be considerably more demanding than reconstruction of low-order correlations. For this reason, we focus on the second-order correlation function $g_{\mathrm{N}}^{(2)}$ as a more robust example of the information that can be extracted from the emission intensity.

Fig.~\ref{fig:reconstruction} illustrates the reconstruction for coherent and thermal number distributions and the dependence of the inferred $g_{\mathrm{N}}^{(2)}$ on $R$, $\eta$, and the mean occupation $\expval{\Nop}$. The known detection efficiency, detector response, and bin integration are included in the forward model, so their effects on the expected signal are accounted for in the reconstruction. This does not eliminate the statistical uncertainty associated with finite detection efficiency or the limitations imposed by finite temporal resolution. Nevertheless, the results demonstrate reconstruction of $g_{\mathrm{N}}^{(2)}$ over the range of efficiencies considered, with increasing numbers of experimental repetitions $R$ reducing the spread across independent reconstructions.

For $R$ independent preparations and complete temporal collection, the expected total number of detected photons is $M_{\mathrm{det}}=\eta R\expval{\Nop}$. Increasing the mean particle number therefore provides a larger photon signal, which can contribute to the improved reconstruction of $g_{\mathrm{N}}^{(2)}$ at larger mean occupations observed in Fig.~\ref{fig:reconstruction}(e). Furthermore, focusing on normalized correlation functions such as $g_{\mathrm{N}}^{(2)}$ facilitates comparison across mean occupations. By contrast, the raw moments $\expval{\Nop^k}$ typically grow as $\expval{\Nop}^k$ at large mean occupation for the distribution families considered here, and can therefore exhibit increasingly large absolute reconstruction errors even at fixed relative accuracy. We also note that with insufficient statistics our estimates are biased, as illustrated for small $\expval{\Nop}$, small $R$, and small $\eta$, where $g_{\mathrm{N}}^{(2)}$ is systematically overestimated. Ultimately, finite temporal resolution limits how far the improvement with increasing particle number can persist: within each fixed-$N$ sector, the characteristic superradiant pulse width decreases as $1/(\Gamma N)$, making sufficiently large-$N$ pulses increasingly difficult to resolve. The cat example in Fig.~\ref{fig:rec_app} illustrates that $g_{\mathrm{N}}^{(2)}$ can remain as robustly reconstructed as for smooth distributions even when the parity structure is unresolved.
 
\section{Atom loss}
\label{sec:atom_loss}

\begin{figure}[!ht]
    \centering
    \includegraphics[width=\linewidth]{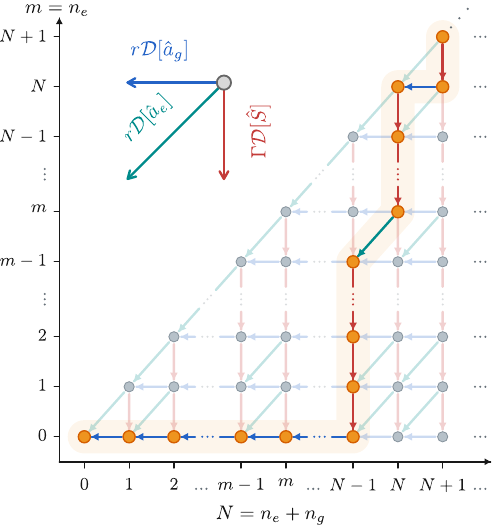}
    \caption{\textbf{Full Dicke triangle with losses.}
    Ground-state loss (blue) and excited-state loss (teal) connect the otherwise disconnected Dicke chains (red).
    A representative relaxation path is highlighted in orange from the fully excited Dicke coordinates $(N+1,N+1)$ to the vacuum coordinates $(0,0)$.
    From state $(N,m)$, ground-state loss, excited-state loss, and collective emission occur at rates $r(N-m)$, $rm$, and $\Gamma h_m^{(N)}$, respectively.
    }
    \label{fig:dicke_triangle}
\end{figure}

The interplay of collective emission with local dissipative processes has been extensively studied using permutation-invariant descriptions~\cite{chase2008,baragiola2010,xu2013,shammah2018}. While these methods substantially reduce the computational complexity, general local dissipative processes can populate states outside the fully symmetric Dicke manifold. Physical atom loss, by contrast, preserves the symmetric manifold of the remaining emitters, allowing the dynamics to be described by transitions between symmetric Dicke manifolds of different particle numbers.

We now include atom loss from both Schwinger-boson modes at the same rate \(r\) as a new Lindblad term
\begin{equation}
    \mathcal{L}_{\mathrm{loss}}\hat{\rho}=r\mathcal{D}[\hat{a}_{\mathrm{e}}]\hat{\rho}+r\mathcal{D}[\hat{a}_{\mathrm{g}}]\hat{\rho}.
\end{equation}

As shown in App.~\ref{app:loss_model}, the Schwinger-boson description of particle loss reproduces the underlying microscopic local-loss model exactly within the particle-number-diagonal density-matrix blocks \(N=N'\). In general, the two descriptions differ in their treatment of coherences between sectors with \(N\neq N'\). However, since the particle-number-diagonal blocks evolve independently of all off-diagonal blocks and the observables considered here conserve total particle number, both descriptions yield identical results for these observables. Note that the equal loss rates for excited- and ground-state atoms we use here for simplicity are essential for the following analytical treatments and in particular our exact results do not follow if the rates differ. Notably, the total particle loss rate would become dependent on the configuration within a fixed particle number sector.

The loss operators do not conserve the total atom number. Instead, they couple the \(N\)-particle sector to the \((N-1)\)-particle sector. For the diagonal populations
\begin{equation}
    p_m^{(N)}(t)=\rho_{m,m}^{(N,N)}(t),
\end{equation}
we obtain the population equations of motion
\begin{equation}\label{eq:loss_pop}
    \begin{aligned}
    \frac{d}{dt}p_m^{(N)}(t)
    =&\Gamma h_{m+1}^{(N)}p_{m+1}^{(N)}(t)
    -\Gamma h_m^{(N)}p_m^{(N)}(t)\\
    &+r(m+1)p_{m+1}^{(N+1)}(t)\\
    &+r(N+1-m)p_m^{(N+1)}(t)
    -rNp_m^{(N)}(t).
    \end{aligned}
\end{equation}

Here $\sum_{m=0}^N p_m^{(N)}(t)=P_N(t)=\Tr\hat\rho^{(N,N)}(t)$, so the sector blocks are unnormalized. The first two terms are the usual Dicke cascade within a fixed \(N\)-sector. The next two terms describe feeding into the \(N\)-particle sector from the \((N+1)\)-particle sector due to the loss of one excited or one ground-state atom. The final term is the total loss out of the \(N\)-particle sector. Since a state with \(N\) atoms contains \(N\) atoms in total, the total loss
rate is \(rN\).

\subsection{Exact dynamics within the initial particle-number sector}
\label{sec:initial_sector}
Consider first the dynamics restricted to the initial $N$-particle column of the Schwinger-boson lattice for an initial Fock state of $N$ particles. The unnormalized density-matrix block obeys
\begin{equation}
    \dot{\hat\rho}^{(N,N)}=\left(\mathcal L_{\mathrm{Dicke}}-rN\mathcal I\right)\hat\rho^{(N,N)},
\end{equation}
where \(\mathcal I\) denotes the identity superoperator and $\mathcal L_{\mathrm{Dicke}}$ is the pure Dicke time evolution Liouvillian. Consequently,
\begin{equation}
    \hat\rho^{(N,N)}(t;r)=\me^{-rNt}e^{\mathcal L_{\mathrm{Dicke}}t}\hat\rho^{(N,N)}(0),
\end{equation}
and for the populations it follows that
\begin{equation}
    p_m^{(N)}(t;r)=\me^{-rNt}\,p_m^{(N)}(t;r=0).
\end{equation}

Thus, on the initial \(N\)-particle column, atom loss only contributes the common survival factor \(\me^{-rNt}\), while the relative Dicke dynamics is unchanged. The lost population feeds the lower particle-number columns.

For the lower $k$-particle sectors, this factorization does not hold anymore. Instead we may consider the marginal $p^{(k)}(t)=\sum_{m}p_m^{(k)}(t)$ whose equation of motion becomes
\begin{equation}
\label{eq:marginal}
\frac{d p^{(k)}}{dt}=r(k+1)p^{(k+1)}-rkp^{(k)},
\end{equation}
which is an exponential decay process solved by \(p^{(k)}(t)=\binom Nk \me^{-kr t}\left(1-e^{-r t}\right)^{N-k}\) for $p^{(k)}(0)=\delta_{k,N}$. Thus, despite the complex dynamics due to the superradiant emission within each $N$-sector and the fact that the joint probability distribution for $(m,N)$ does not factorize, the particle loss is not affected by Dicke superradiance so that the marginal gives the fully analytically solvable particle loss Markov chain for $N$.

\subsection{Large-N scaling functions}
\label{sec:large_n_scaling}

\begin{figure}
    \centering
    \includegraphics[width=\columnwidth]{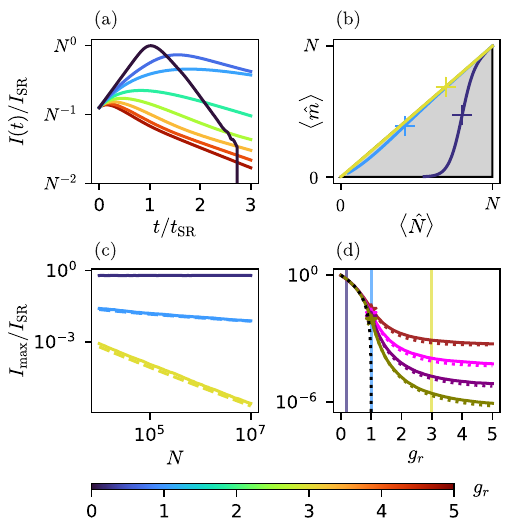}
    \caption{\textbf{Atom loss and peak-intensity scaling.} (a) Emission profiles for \(N=10^7\) at several values of \(g_{\mathrm{r}}\), indicated by the colorbar.
        Intensity is normalized to the loss-free peak \(I_{\mathrm{SR}}\) and time to \(t_{\mathrm{SR}}\). Monte Carlo averages use $n_{\mathrm{T}}=10^3$ trajectories for $g_{\mathrm{r}}>0$ and $10^4$ for $g_{\mathrm{r}}=0$.
        (b) Selected corresponding trajectories on the Dicke triangle, parametrized by \(\langle\hat N\rangle\) and \(\langle\hat m\rangle\) for initial particle number $N_0=500$, with matching colors, obtained by direct integration of Eq.~\eqref{eq:loss_pop}.
        (c) Peak intensity versus \(N\) for fixed \(g_{\mathrm{r}}=0.2,1,3\), illustrating quadratic, critical, and logarithmically corrected subquadratic scaling, respectively, using $n_{\mathrm{T}}=10^3$.
        (d) Normalized peak intensity versus \(g_{\mathrm{r}}\) for \(N=10^4,10^5,10^6,10^7\). Monte Carlo averages use $n_{\mathrm{T}}=10^4$ for $N=10^4,10^5$ and $10^3$ for $N=10^6,10^7$, except for the loss-free $N=10^7$ point, which uses $10^4$. Dotted lines show the large-\(N\) analytical predictions, while the vertical lines identify the values used in (c).}
    \label{fig:fig6}
\end{figure}

Having established the exact dynamics of particle loss, we now investigate its effect on the superradiant peak in the large-\(N\) limit. Previous studies have investigated Dicke superradiance in the presence of local dephasing~\cite{shammah2017}, its large-\(N\) dynamics with and without incoherent excitation decay~\cite{malz2022}, and the scaling consequences of competing decoherence processes~\cite{bassler2026}. Physical atom loss and replenishment have also been considered in the context of superradiant lasing, including transient emission~\cite{zhang2018}. Here, we investigate how the competition between collective emission and physical removal of excited- and ground-state atoms determines the scaling of the superradiant peak intensity. The exact initial intensity slope is
\begin{equation}
    \frac{\dot I(0)}{\Gamma^2}=\expval{\Nop^2}-\left(2+\frac r\Gamma\right)\expval{\Nop}
\end{equation}
so, for fixed initial particle number $N$, the intensity initially increases if and only if $\Gamma(N-2)>r$, reducing to $N\Gamma>r$ at leading order for $N\gg1$. For an initial number distribution with nonzero mean, the exact condition generalizes to
\begin{equation}
    \expval{\Nop}g_{\mathrm{N}}^{(2)}>1+\frac r\Gamma,
\end{equation}
which extends the loss-free criterion in Sec.~\ref{sec:initial_statistics}. However, initial growth does not determine whether the intensity scales as $I_{\mathrm{max}}\propto N^2$ or as some other function of $N$. We can determine the correct scaling of the emitter loss and superradiant emission rate by inspecting their respective time scales

\begin{equation}
t_{\mathrm{SR}}=\frac{\ln N}{\Gamma N},
\qquad
t_{\mathrm{loss}}=\frac1r,
\end{equation}
whose ratio we define as the dimensionless scaling variable
\begin{equation}
    g_{\mathrm{r}}=\frac{t_{\mathrm{SR}}}{t_{\mathrm{loss}}}=\frac{r\ln N}{\Gamma N}.
\end{equation}

Keeping $g_{\mathrm{r}}$ fixed, we first analyze the mean-field equations of motion, factorizing $\expval{\Nop\mop}\approx \expval{\Nop}\expval{\mop}$ and $\expval{\mop^2}\approx\expval{\mop}^2$ for the operators defined in Eq.~\eqref{eq:number_operators}, as detailed in App.~\ref{app:mean_field}. Including the stochastic delay distribution as in App.~\ref{app:stochastic_asymptotics} yields the peak intensity of the ensemble-averaged trace in the limit $N\to\infty$ at fixed $g_{\mathrm{r}}$, with $r=g_{\mathrm{r}}\Gamma N/\ln N$:
\begin{equation}\label{eq:scaling_laws}
    I_{\mathrm{max}}\sim
\begin{cases}
\displaystyle0.1957\Gamma N^2(1-g_{\mathrm{r}})^2,&0\leq g_{\mathrm{r}}<1,\\
\displaystyle0.3891\frac{\Gamma N^2}{\ln^2N},&g_{\mathrm{r}}=1,\\
\displaystyle\frac{4g_{\mathrm{r}}^2}{e^2}\frac{\Gamma N^{1+1/g_{\mathrm{r}}}}{\ln^2N},&g_{\mathrm{r}}>1.
\end{cases}
\end{equation}

These are pointwise asymptotics at fixed $g_{\mathrm{r}}$. The outer branches are not uniform within $|g_{\mathrm{r}}-1|=O(1/\ln N)$. In particular, the $g_{\mathrm{r}}>1$ branch does not extend to $g_{\mathrm{r}}=O(\ln N)$, where $\ln N/g_{\mathrm{r}}$ need not be large. For comparison, the mean-field prefactors are $(1-g_{\mathrm{r}})^2/4$ and $0.4392$ in the first two regimes; the leading loss-dominated result is unchanged by delay averaging.

We therefore distinguish three scaling regimes, separated by the critical point \(g_{\mathrm{r}}=1\) and a crossover to independent emission at sufficiently strong loss:
\begin{enumerate}
    \item $g_{\mathrm{r}}<1\Leftrightarrow r\ln N<\Gamma N$: Superradiant regime with $I_{\mathrm{max}}\propto N^2$.
    \item Intermediate regime for $1<g_{\mathrm{r}}\ll\ln N$, with $I_{\mathrm{max}}\propto \frac{\Gamma N^{1+1/g_{\mathrm{r}}}}{\ln^2N}$.
    \item $N\Gamma<r$: Linear regime where $I_{\mathrm{max}}\propto N$.
\end{enumerate}

We illustrate the intensity dynamics and the three scaling regimes in Fig.~\ref{fig:fig6}. We obtain the intensity traces and peak intensities in Fig.~\ref{fig:fig6}(a), (c), and (d) from Gillespie Monte Carlo simulation~\cite{gillespie1976general} of the population equations of motion, estimating the intensity as
\begin{equation}
   \overline I(t)=\frac{\Gamma}{n_{\mathrm{T}}}\sum_{s=1}^{n_{\mathrm{T}}}h_{m_s(t)}^{(N_s(t))},
\end{equation}
where $n_{\mathrm{T}}$ is the number of independent trajectories and \(N_s(t)\) and \(m_s(t)\) denote the particle and excitation numbers along trajectory \(s\). Peak intensities are extracted as \(I_{\mathrm{max}}=\max_t\overline I(t)\).

First, we show ensemble-averaged intensity traces in Fig.~\ref{fig:fig6}(a) and their corresponding trajectories on the Dicke triangle in Fig.~\ref{fig:fig6}(b), where we plot $\expval{\hat N}$ and $\expval{\hat m}$ traces. Increasing particle loss suppresses the superradiant peak and causes the trajectories to penetrate further into the interior of the triangle. The trajectories also illustrate an approximate separation of the dynamics into three stages: first, a fraction of the excited emitters is lost; second, the remaining emitters undergo collective radiative decay; and finally, the residual ground-state population is lost. Since collective emission conserves the particle number, the second stage corresponds to an approximately vertical trajectory in the Dicke triangle, particularly at large $N$.

In Fig.~\ref{fig:fig6}(c), we investigate the scaling of the peak intensity with $N$ for several fixed values of $g_{\mathrm{r}}$, marked by vertical lines in Fig.~\ref{fig:fig6}(d). The numerical results confirm the scaling laws in Eq.~\eqref{eq:scaling_laws}, with the largest finite-size corrections occurring near $g_{\mathrm{r}}=1$. Finally, Fig.~\ref{fig:fig6}(d) compares the peak intensity, normalized by the loss-free Dicke peak $I_{\mathrm{SR}}\approx\Gamma N^2/5$, with the analytical asymptotic expressions as a function of $g_{\mathrm{r}}$ for different system sizes. We find good agreement away from $g_{\mathrm{r}}=1$, whereas convergence is considerably slower near the critical point. In particular, the separate asymptotic branches do not connect continuously at finite $N$, although the numerical peak intensity is continuous in $g_{\mathrm{r}}$. This reflects the nonuniformity of the asymptotic expansion near $g_{\mathrm{r}}=1$, where the crossover region shrinks only as $1/\ln N$.

Combining loss with initial particle distributions is straightforward via the relation
\begin{equation}
    I(t)=\sum_N P_N I_N(t;r),
\end{equation}
where $I_N(t;r)$ is the full intensity trace with fixed atom loss $r$ and decay rate $\Gamma$ for initial particle number $N$. The kernels $I_N(t;r)$ can be calculated from Eq.~\eqref{eq:loss_pop} and thus the treatment of initial number fluctuations is straightforward. However, we do not pursue this here in detail. We also have not included loss or decoherence in the reconstruction; such channels may reduce its fidelity.

\section{Conclusions}
\label{sec:conclusions}
In summary, we have investigated Dicke superradiance in ensembles with variable emitter number, arising from both initial number fluctuations and continuous particle loss. The Schwinger-boson representation provides a unified framework for these situations. In the absence of loss, the emission intensity of an initially number-indefinite ensemble reduces to a weighted collection of independent fixed-$N$ Dicke cascades, as expressed in Eq.~\eqref{eq:number_averaged_intensity}.

We have shown that initial number fluctuations directly influence the superradiant emission dynamics. For initially fully excited ensembles, the criterion for an initially increasing intensity depends on both the mean particle number and its fluctuations, $\langle\Nop\rangle g_{\mathrm{N}}^{(2)}>1$. Consequently, sufficiently broad number distributions can exhibit initial intensity growth even at small mean occupations. Conversely, the different temporal profiles of the individual number sectors encode information about the initial number distribution in the ensemble-averaged intensity. We have demonstrated that the second-order atom-number correlation $g_{\mathrm{N}}^{(2)}$ can be reconstructed from finite-statistics intensity measurements, even when the detailed number distribution remains difficult to resolve.

For continuous particle loss, we have established the equivalence between the Schwinger-boson description and a microscopic model of local atom loss for the relevant particle-number-diagonal dynamics. With equal loss rates from the two internal states, the populations remaining within the initial particle-number sector follow the loss-free Dicke dynamics multiplied by the survival factor $\me^{-rNt}$, while the total particle-number distribution evolves according to binomial loss statistics independently of collective emission.

Finally, we have investigated the competition between collective emission and particle loss in the large-$N$ limit. The ratio of the superradiant buildup and particle-loss timescales defines the scaling variable $g_{\mathrm{r}}=r\ln N/(\Gamma N)$. Depending on this ratio, the peak intensity exhibits quadratic superradiant enhancement, logarithmically corrected subquadratic power laws, or crosses over to linear scaling under sufficiently strong loss. The initial-growth condition and the asymptotic scaling of the peak intensity therefore characterize distinct aspects of the collective emission dynamics.

Taken together, these results show how variable emitter number influences both the temporal signatures of Dicke superradiance and the collective enhancement that survives in the presence of particle loss.

\vspace*{0.5cm}
\noindent \textbf{Acknowledgments---} The authors acknowledge the use of OpenAI tools such as ChatGPT (GPT-5 and GPT-6) and Codex to assist with code generation, conceptual figure design and preparation, literature searches, and language editing. All AI-assisted contributions were critically evaluated and independently verified by the authors, who take full responsibility for the scientific content of this work.\\

\noindent \textbf{Code and data availability.}
The simulation code, plotting scripts, and reduced data used to generate the figures are available in the accompanying GitHub repository~\cite{dicke_superradiance_reproducibility}.

\bibliography{Refs}

\begin{thebibliography}{44}%
\makeatletter
\providecommand \@ifxundefined [1]{%
 \@ifx{#1\undefined}
}%
\providecommand \@ifnum [1]{%
 \ifnum #1\expandafter \@firstoftwo
 \else \expandafter \@secondoftwo
 \fi
}%
\providecommand \@ifx [1]{%
 \ifx #1\expandafter \@firstoftwo
 \else \expandafter \@secondoftwo
 \fi
}%
\providecommand \natexlab [1]{#1}%
\providecommand \enquote  [1]{``#1''}%
\providecommand \bibnamefont  [1]{#1}%
\providecommand \bibfnamefont [1]{#1}%
\providecommand \citenamefont [1]{#1}%
\providecommand \href@noop [0]{\@secondoftwo}%
\providecommand \href [0]{\begingroup \@sanitize@url \@href}%
\providecommand \@href[1]{\@@startlink{#1}\@@href}%
\providecommand \@@href[1]{\endgroup#1\@@endlink}%
\providecommand \@sanitize@url [0]{\catcode `\\12\catcode `\$12\catcode
  `\&12\catcode `\#12\catcode `\^12\catcode `\_12\catcode `\%12\relax}%
\providecommand \@@startlink[1]{}%
\providecommand \@@endlink[0]{}%
\providecommand \url  [0]{\begingroup\@sanitize@url \@url }%
\providecommand \@url [1]{\endgroup\@href {#1}{\urlprefix }}%
\providecommand \urlprefix  [0]{URL }%
\providecommand \Eprint [0]{\href }%
\providecommand \doibase [0]{https://doi.org/}%
\providecommand \selectlanguage [0]{\@gobble}%
\providecommand \bibinfo  [0]{\@secondoftwo}%
\providecommand \bibfield  [0]{\@secondoftwo}%
\providecommand \translation [1]{[#1]}%
\providecommand \BibitemOpen [0]{}%
\providecommand \bibitemStop [0]{}%
\providecommand \bibitemNoStop [0]{.\EOS\space}%
\providecommand \EOS [0]{\spacefactor3000\relax}%
\providecommand \BibitemShut  [1]{\csname bibitem#1\endcsname}%
\let\auto@bib@innerbib\@empty
\bibitem [{\citenamefont {Dicke}(1954)}]{dicke1954}%
  \BibitemOpen
  \bibfield  {author} {\bibinfo {author} {\bibfnamefont {R.~H.}\ \bibnamefont
  {Dicke}},\ }\bibfield  {title} {\bibinfo {title} {Coherence in spontaneous
  radiation processes},\ }\href {https://doi.org/10.1103/PhysRev.93.99}
  {\bibfield  {journal} {\bibinfo  {journal} {Phys. Rev.}\ }\textbf {\bibinfo
  {volume} {93}},\ \bibinfo {pages} {99} (\bibinfo {year} {1954})}\BibitemShut
  {NoStop}%
\bibitem [{\citenamefont {Gross}\ and\ \citenamefont
  {Haroche}(1982)}]{gross1982}%
  \BibitemOpen
  \bibfield  {author} {\bibinfo {author} {\bibfnamefont {M.}~\bibnamefont
  {Gross}}\ and\ \bibinfo {author} {\bibfnamefont {S.}~\bibnamefont
  {Haroche}},\ }\bibfield  {title} {\bibinfo {title} {Superradiance: An essay
  on the theory of collective spontaneous emission},\ }\href
  {https://doi.org/10.1016/0370-1573(82)90102-8} {\bibfield  {journal}
  {\bibinfo  {journal} {Physics Reports}\ }\textbf {\bibinfo {volume} {93}},\
  \bibinfo {pages} {301} (\bibinfo {year} {1982})}\BibitemShut {NoStop}%
\bibitem [{\citenamefont {Norcia}\ \emph {et~al.}(2016)\citenamefont {Norcia},
  \citenamefont {Winchester}, \citenamefont {Cline},\ and\ \citenamefont
  {Thompson}}]{norcia20162}%
  \BibitemOpen
  \bibfield  {author} {\bibinfo {author} {\bibfnamefont {M.~A.}\ \bibnamefont
  {Norcia}}, \bibinfo {author} {\bibfnamefont {M.~N.}\ \bibnamefont
  {Winchester}}, \bibinfo {author} {\bibfnamefont {J.~R.~K.}\ \bibnamefont
  {Cline}},\ and\ \bibinfo {author} {\bibfnamefont {J.~K.}\ \bibnamefont
  {Thompson}},\ }\bibfield  {title} {\bibinfo {title} {Superradiance on the
  millihertz linewidth strontium clock transition},\ }\href
  {https://doi.org/10.1126/sciadv.1601231} {\bibfield  {journal} {\bibinfo
  {journal} {Science Advances}\ }\textbf {\bibinfo {volume} {2}},\ \bibinfo
  {pages} {e1601231} (\bibinfo {year} {2016})}\BibitemShut {NoStop}%
\bibitem [{\citenamefont {Liedl}\ \emph {et~al.}(2024)\citenamefont {Liedl},
  \citenamefont {Tebbenjohanns}, \citenamefont {Bach}, \citenamefont {Pucher},
  \citenamefont {Rauschenbeutel},\ and\ \citenamefont
  {Schneeweiss}}]{liedl2024}%
  \BibitemOpen
  \bibfield  {author} {\bibinfo {author} {\bibfnamefont {C.}~\bibnamefont
  {Liedl}}, \bibinfo {author} {\bibfnamefont {F.}~\bibnamefont
  {Tebbenjohanns}}, \bibinfo {author} {\bibfnamefont {C.}~\bibnamefont {Bach}},
  \bibinfo {author} {\bibfnamefont {S.}~\bibnamefont {Pucher}}, \bibinfo
  {author} {\bibfnamefont {A.}~\bibnamefont {Rauschenbeutel}},\ and\ \bibinfo
  {author} {\bibfnamefont {P.}~\bibnamefont {Schneeweiss}},\ }\bibfield
  {title} {\bibinfo {title} {Observation of superradiant bursts in a cascaded
  quantum system},\ }\href {https://doi.org/10.1103/PhysRevX.14.011020}
  {\bibfield  {journal} {\bibinfo  {journal} {Physical Review X}\ }\textbf
  {\bibinfo {volume} {14}},\ \bibinfo {pages} {011020} (\bibinfo {year}
  {2024})}\BibitemShut {NoStop}%
\bibitem [{\citenamefont {Ferioli}\ \emph {et~al.}(2025)\citenamefont
  {Ferioli}, \citenamefont {Ferrier-Barbut},\ and\ \citenamefont
  {Browaeys}}]{ferioli2025}%
  \BibitemOpen
  \bibfield  {author} {\bibinfo {author} {\bibfnamefont {G.}~\bibnamefont
  {Ferioli}}, \bibinfo {author} {\bibfnamefont {I.}~\bibnamefont
  {Ferrier-Barbut}},\ and\ \bibinfo {author} {\bibfnamefont {A.}~\bibnamefont
  {Browaeys}},\ }\bibfield  {title} {\bibinfo {title} {Emergence of
  second-order coherence in the superradiant emission from a free-space atomic
  ensemble},\ }\href {https://doi.org/10.1103/PhysRevLett.134.153602}
  {\bibfield  {journal} {\bibinfo  {journal} {Phys. Rev. Lett.}\ }\textbf
  {\bibinfo {volume} {134}},\ \bibinfo {pages} {153602} (\bibinfo {year}
  {2025})}\BibitemShut {NoStop}%
\bibitem [{\citenamefont {Bach}\ \emph {et~al.}(2026)\citenamefont {Bach},
  \citenamefont {Tebbenjohanns}, \citenamefont {Liedl}, \citenamefont
  {Schneeweiss},\ and\ \citenamefont {Rauschenbeutel}}]{bach2026emergence}%
  \BibitemOpen
  \bibfield  {author} {\bibinfo {author} {\bibfnamefont {C.}~\bibnamefont
  {Bach}}, \bibinfo {author} {\bibfnamefont {F.}~\bibnamefont {Tebbenjohanns}},
  \bibinfo {author} {\bibfnamefont {C.}~\bibnamefont {Liedl}}, \bibinfo
  {author} {\bibfnamefont {P.}~\bibnamefont {Schneeweiss}},\ and\ \bibinfo
  {author} {\bibfnamefont {A.}~\bibnamefont {Rauschenbeutel}},\ }\bibfield
  {title} {\bibinfo {title} {Emergence of second-order coherence in
  superfluorescence},\ }\href {https://doi.org/10.1103/r3kg-c4x9} {\bibfield
  {journal} {\bibinfo  {journal} {Phys. Rev. Lett.}\ }\textbf {\bibinfo
  {volume} {136}},\ \bibinfo {pages} {063402} (\bibinfo {year}
  {2026})}\BibitemShut {NoStop}%
\bibitem [{\citenamefont {Orzel}\ \emph {et~al.}(2001)\citenamefont {Orzel},
  \citenamefont {Tuchman}, \citenamefont {Fenselau}, \citenamefont {Yasuda},\
  and\ \citenamefont {Kasevich}}]{orzel2001}%
  \BibitemOpen
  \bibfield  {author} {\bibinfo {author} {\bibfnamefont {C.}~\bibnamefont
  {Orzel}}, \bibinfo {author} {\bibfnamefont {A.~K.}\ \bibnamefont {Tuchman}},
  \bibinfo {author} {\bibfnamefont {M.~L.}\ \bibnamefont {Fenselau}}, \bibinfo
  {author} {\bibfnamefont {M.}~\bibnamefont {Yasuda}},\ and\ \bibinfo {author}
  {\bibfnamefont {M.~A.}\ \bibnamefont {Kasevich}},\ }\bibfield  {title}
  {\bibinfo {title} {Squeezed states in a {Bose}-{Einstein} condensate},\
  }\href {https://doi.org/10.1126/science.1058149} {\bibfield  {journal}
  {\bibinfo  {journal} {Science}\ }\textbf {\bibinfo {volume} {291}},\ \bibinfo
  {pages} {2386} (\bibinfo {year} {2001})}\BibitemShut {NoStop}%
\bibitem [{\citenamefont {Kristensen}\ \emph {et~al.}(2019)\citenamefont
  {Kristensen}, \citenamefont {Christensen}, \citenamefont {Gajdacz},
  \citenamefont {Iglicki}, \citenamefont {Paw\l{}owski}, \citenamefont
  {Klempt}, \citenamefont {Sherson}, \citenamefont {Rz{\k{a}}{\.{z}}ewski},
  \citenamefont {Hilliard},\ and\ \citenamefont {Arlt}}]{kristensen2019}%
  \BibitemOpen
  \bibfield  {author} {\bibinfo {author} {\bibfnamefont {M.~A.}\ \bibnamefont
  {Kristensen}}, \bibinfo {author} {\bibfnamefont {M.~B.}\ \bibnamefont
  {Christensen}}, \bibinfo {author} {\bibfnamefont {M.}~\bibnamefont
  {Gajdacz}}, \bibinfo {author} {\bibfnamefont {M.}~\bibnamefont {Iglicki}},
  \bibinfo {author} {\bibfnamefont {K.}~\bibnamefont {Paw\l{}owski}}, \bibinfo
  {author} {\bibfnamefont {C.}~\bibnamefont {Klempt}}, \bibinfo {author}
  {\bibfnamefont {J.~F.}\ \bibnamefont {Sherson}}, \bibinfo {author}
  {\bibfnamefont {K.}~\bibnamefont {Rz{\k{a}}{\.{z}}ewski}}, \bibinfo {author}
  {\bibfnamefont {A.~J.}\ \bibnamefont {Hilliard}},\ and\ \bibinfo {author}
  {\bibfnamefont {J.~J.}\ \bibnamefont {Arlt}},\ }\bibfield  {title} {\bibinfo
  {title} {Observation of atom number fluctuations in a {Bose}-{Einstein}
  condensate},\ }\href {https://doi.org/10.1103/PhysRevLett.122.163601}
  {\bibfield  {journal} {\bibinfo  {journal} {Phys. Rev. Lett.}\ }\textbf
  {\bibinfo {volume} {122}},\ \bibinfo {pages} {163601} (\bibinfo {year}
  {2019})}\BibitemShut {NoStop}%
\bibitem [{\citenamefont {Kruk}\ \emph {et~al.}(2025)\citenamefont {Kruk},
  \citenamefont {Kulik}, \citenamefont {Andersen}, \citenamefont {Deuar},
  \citenamefont {Gajda}, \citenamefont {Paw{\l}owski}, \citenamefont
  {Witkowska}, \citenamefont {Arlt},\ and\ \citenamefont
  {Rz{\k{a}}{\.{z}}ewski}}]{kruk2025}%
  \BibitemOpen
  \bibfield  {author} {\bibinfo {author} {\bibfnamefont {M.~B.}\ \bibnamefont
  {Kruk}}, \bibinfo {author} {\bibfnamefont {P.}~\bibnamefont {Kulik}},
  \bibinfo {author} {\bibfnamefont {M.~F.}\ \bibnamefont {Andersen}}, \bibinfo
  {author} {\bibfnamefont {P.}~\bibnamefont {Deuar}}, \bibinfo {author}
  {\bibfnamefont {M.}~\bibnamefont {Gajda}}, \bibinfo {author} {\bibfnamefont
  {K.}~\bibnamefont {Paw{\l}owski}}, \bibinfo {author} {\bibfnamefont
  {E.}~\bibnamefont {Witkowska}}, \bibinfo {author} {\bibfnamefont {J.~J.}\
  \bibnamefont {Arlt}},\ and\ \bibinfo {author} {\bibfnamefont
  {K.}~\bibnamefont {Rz{\k{a}}{\.{z}}ewski}},\ }\bibfield  {title} {\bibinfo
  {title} {On the fluctuations of the number of atoms in the condensate},\
  }\href {https://doi.org/10.1088/1361-6633/ae0e33} {\bibfield  {journal}
  {\bibinfo  {journal} {Reports on Progress in Physics}\ }\textbf {\bibinfo
  {volume} {88}},\ \bibinfo {pages} {106401} (\bibinfo {year}
  {2025})}\BibitemShut {NoStop}%
\bibitem [{\citenamefont {Goban}\ \emph {et~al.}(2015)\citenamefont {Goban},
  \citenamefont {Hung}, \citenamefont {Hood}, \citenamefont {Yu}, \citenamefont
  {Muniz}, \citenamefont {Painter},\ and\ \citenamefont {Kimble}}]{goban2015}%
  \BibitemOpen
  \bibfield  {author} {\bibinfo {author} {\bibfnamefont {A.}~\bibnamefont
  {Goban}}, \bibinfo {author} {\bibfnamefont {C.-L.}\ \bibnamefont {Hung}},
  \bibinfo {author} {\bibfnamefont {J.~D.}\ \bibnamefont {Hood}}, \bibinfo
  {author} {\bibfnamefont {S.-P.}\ \bibnamefont {Yu}}, \bibinfo {author}
  {\bibfnamefont {J.~A.}\ \bibnamefont {Muniz}}, \bibinfo {author}
  {\bibfnamefont {O.}~\bibnamefont {Painter}},\ and\ \bibinfo {author}
  {\bibfnamefont {H.~J.}\ \bibnamefont {Kimble}},\ }\bibfield  {title}
  {\bibinfo {title} {Superradiance for atoms trapped along a photonic crystal
  waveguide},\ }\href {https://doi.org/10.1103/PhysRevLett.115.063601}
  {\bibfield  {journal} {\bibinfo  {journal} {Phys. Rev. Lett.}\ }\textbf
  {\bibinfo {volume} {115}},\ \bibinfo {pages} {063601} (\bibinfo {year}
  {2015})}\BibitemShut {NoStop}%
\bibitem [{\citenamefont {Robicheaux}(2021)}]{robicheaux20212}%
  \BibitemOpen
  \bibfield  {author} {\bibinfo {author} {\bibfnamefont {F.}~\bibnamefont
  {Robicheaux}},\ }\bibfield  {title} {\bibinfo {title} {Theoretical study of
  early-time superradiance for atom clouds and arrays},\ }\href
  {https://doi.org/10.1103/PhysRevA.104.063706} {\bibfield  {journal} {\bibinfo
   {journal} {Phys. Rev. A}\ }\textbf {\bibinfo {volume} {104}},\ \bibinfo
  {pages} {063706} (\bibinfo {year} {2021})}\BibitemShut {NoStop}%
\bibitem [{\citenamefont {Masson}\ and\ \citenamefont
  {Asenjo-Garcia}(2022)}]{masson2022}%
  \BibitemOpen
  \bibfield  {author} {\bibinfo {author} {\bibfnamefont {S.~J.}\ \bibnamefont
  {Masson}}\ and\ \bibinfo {author} {\bibfnamefont {A.}~\bibnamefont
  {Asenjo-Garcia}},\ }\bibfield  {title} {\bibinfo {title} {Universality of
  {Dicke} superradiance in arrays of quantum emitters},\ }\href
  {https://doi.org/10.1038/s41467-022-29805-4} {\bibfield  {journal} {\bibinfo
  {journal} {Nature Communications}\ }\textbf {\bibinfo {volume} {13}},\
  \bibinfo {pages} {2285} (\bibinfo {year} {2022})}\BibitemShut {NoStop}%
\bibitem [{\citenamefont {Mekhov}\ \emph {et~al.}(2007)\citenamefont {Mekhov},
  \citenamefont {Maschler},\ and\ \citenamefont {Ritsch}}]{mekhov2007}%
  \BibitemOpen
  \bibfield  {author} {\bibinfo {author} {\bibfnamefont {I.~B.}\ \bibnamefont
  {Mekhov}}, \bibinfo {author} {\bibfnamefont {C.}~\bibnamefont {Maschler}},\
  and\ \bibinfo {author} {\bibfnamefont {H.}~\bibnamefont {Ritsch}},\
  }\bibfield  {title} {\bibinfo {title} {Cavity-enhanced light scattering in
  optical lattices to probe atomic quantum statistics},\ }\href
  {https://doi.org/10.1103/PhysRevLett.98.100402} {\bibfield  {journal}
  {\bibinfo  {journal} {Phys. Rev. Lett.}\ }\textbf {\bibinfo {volume} {98}},\
  \bibinfo {pages} {100402} (\bibinfo {year} {2007})}\BibitemShut {NoStop}%
\bibitem [{\citenamefont {Belliardo}\ \emph {et~al.}(2026)\citenamefont
  {Belliardo}, \citenamefont {Chu}, \citenamefont {Koppenh\"{o}fer},\ and\
  \citenamefont {Clerk}}]{belliardo2026}%
  \BibitemOpen
  \bibfield  {author} {\bibinfo {author} {\bibfnamefont {F.}~\bibnamefont
  {Belliardo}}, \bibinfo {author} {\bibfnamefont {A.}~\bibnamefont {Chu}},
  \bibinfo {author} {\bibfnamefont {M.}~\bibnamefont {Koppenh\"{o}fer}},\ and\
  \bibinfo {author} {\bibfnamefont {A.~A.}\ \bibnamefont {Clerk}},\ }\href
  {https://arxiv.org/abs/2603.13130} {\bibinfo {title} {Extracting information
  from a superradiant burst using simple measurements}} (\bibinfo {year}
  {2026}),\ \Eprint {https://arxiv.org/abs/2603.13130} {arXiv:2603.13130
  [quant-ph]} \BibitemShut {NoStop}%
\bibitem [{\citenamefont {Shammah}\ \emph {et~al.}(2017)\citenamefont
  {Shammah}, \citenamefont {Lambert}, \citenamefont {Nori},\ and\ \citenamefont
  {De~Liberato}}]{shammah2017}%
  \BibitemOpen
  \bibfield  {author} {\bibinfo {author} {\bibfnamefont {N.}~\bibnamefont
  {Shammah}}, \bibinfo {author} {\bibfnamefont {N.}~\bibnamefont {Lambert}},
  \bibinfo {author} {\bibfnamefont {F.}~\bibnamefont {Nori}},\ and\ \bibinfo
  {author} {\bibfnamefont {S.}~\bibnamefont {De~Liberato}},\ }\bibfield
  {title} {\bibinfo {title} {Superradiance with local phase-breaking effects},\
  }\href {https://doi.org/10.1103/PhysRevA.96.023863} {\bibfield  {journal}
  {\bibinfo  {journal} {Phys. Rev. A}\ }\textbf {\bibinfo {volume} {96}},\
  \bibinfo {pages} {023863} (\bibinfo {year} {2017})}\BibitemShut {NoStop}%
\bibitem [{\citenamefont {Bassler}\ \emph {et~al.}(2026)\citenamefont
  {Bassler}, \citenamefont {Lyne},\ and\ \citenamefont {Cuerda}}]{bassler2026}%
  \BibitemOpen
  \bibfield  {author} {\bibinfo {author} {\bibfnamefont {N.~S.}\ \bibnamefont
  {Bassler}}, \bibinfo {author} {\bibfnamefont {J.}~\bibnamefont {Lyne}},\ and\
  \bibinfo {author} {\bibfnamefont {J.}~\bibnamefont {Cuerda}},\ }\href
  {https://arxiv.org/abs/2607.28034} {\bibinfo {title} {Scaling theory of
  decoherence in {Dicke} superradiance}} (\bibinfo {year} {2026}),\ \Eprint
  {https://arxiv.org/abs/2607.28034} {arXiv:2607.28034 [quant-ph]} \BibitemShut
  {NoStop}%
\bibitem [{\citenamefont {Zhang}\ \emph {et~al.}(2018)\citenamefont {Zhang},
  \citenamefont {Zhang},\ and\ \citenamefont {M{\o}lmer}}]{zhang2018}%
  \BibitemOpen
  \bibfield  {author} {\bibinfo {author} {\bibfnamefont {Y.}~\bibnamefont
  {Zhang}}, \bibinfo {author} {\bibfnamefont {Y.-X.}\ \bibnamefont {Zhang}},\
  and\ \bibinfo {author} {\bibfnamefont {K.}~\bibnamefont {M{\o}lmer}},\
  }\bibfield  {title} {\bibinfo {title} {{Monte}-{Carlo} simulations of
  superradiant lasing},\ }\href {https://doi.org/10.1088/1367-2630/aaec36}
  {\bibfield  {journal} {\bibinfo  {journal} {New Journal of Physics}\ }\textbf
  {\bibinfo {volume} {20}},\ \bibinfo {pages} {112001} (\bibinfo {year}
  {2018})}\BibitemShut {NoStop}%
\bibitem [{\citenamefont {Bonifacio}\ \emph
  {et~al.}(1971{\natexlab{a}})\citenamefont {Bonifacio}, \citenamefont
  {Schwendimann},\ and\ \citenamefont {Haake}}]{bonifacio19711}%
  \BibitemOpen
  \bibfield  {author} {\bibinfo {author} {\bibfnamefont {R.}~\bibnamefont
  {Bonifacio}}, \bibinfo {author} {\bibfnamefont {P.}~\bibnamefont
  {Schwendimann}},\ and\ \bibinfo {author} {\bibfnamefont {F.}~\bibnamefont
  {Haake}},\ }\bibfield  {title} {\bibinfo {title} {Quantum statistical theory
  of superradiance. {I}},\ }\href {https://doi.org/10.1103/PhysRevA.4.302}
  {\bibfield  {journal} {\bibinfo  {journal} {Phys. Rev. A}\ }\textbf {\bibinfo
  {volume} {4}},\ \bibinfo {pages} {302} (\bibinfo {year}
  {1971}{\natexlab{a}})}\BibitemShut {NoStop}%
\bibitem [{\citenamefont {Bonifacio}\ \emph
  {et~al.}(1971{\natexlab{b}})\citenamefont {Bonifacio}, \citenamefont
  {Schwendimann},\ and\ \citenamefont {Haake}}]{bonifacio19712}%
  \BibitemOpen
  \bibfield  {author} {\bibinfo {author} {\bibfnamefont {R.}~\bibnamefont
  {Bonifacio}}, \bibinfo {author} {\bibfnamefont {P.}~\bibnamefont
  {Schwendimann}},\ and\ \bibinfo {author} {\bibfnamefont {F.}~\bibnamefont
  {Haake}},\ }\bibfield  {title} {\bibinfo {title} {Quantum statistical theory
  of superradiance. {II}},\ }\href {https://doi.org/10.1103/PhysRevA.4.854}
  {\bibfield  {journal} {\bibinfo  {journal} {Phys. Rev. A}\ }\textbf {\bibinfo
  {volume} {4}},\ \bibinfo {pages} {854} (\bibinfo {year}
  {1971}{\natexlab{b}})}\BibitemShut {NoStop}%
\bibitem [{\citenamefont {Haake}\ and\ \citenamefont
  {Glauber}(1972)}]{haake1972}%
  \BibitemOpen
  \bibfield  {author} {\bibinfo {author} {\bibfnamefont {F.}~\bibnamefont
  {Haake}}\ and\ \bibinfo {author} {\bibfnamefont {R.~J.}\ \bibnamefont
  {Glauber}},\ }\bibfield  {title} {\bibinfo {title} {Quantum statistics of
  superradiant pulses},\ }\href {https://doi.org/10.1103/PhysRevA.5.1457}
  {\bibfield  {journal} {\bibinfo  {journal} {Phys. Rev. A}\ }\textbf {\bibinfo
  {volume} {5}},\ \bibinfo {pages} {1457} (\bibinfo {year} {1972})}\BibitemShut
  {NoStop}%
\bibitem [{\citenamefont {Degiorgio}(1971)}]{degiorgio19711}%
  \BibitemOpen
  \bibfield  {author} {\bibinfo {author} {\bibfnamefont {V.}~\bibnamefont
  {Degiorgio}},\ }\bibfield  {title} {\bibinfo {title} {Statistical properties
  of superradiant pulses},\ }\href
  {https://doi.org/10.1016/0030-4018(71)90245-8} {\bibfield  {journal}
  {\bibinfo  {journal} {Optics Communications}\ }\textbf {\bibinfo {volume}
  {2}},\ \bibinfo {pages} {362} (\bibinfo {year} {1971})}\BibitemShut {NoStop}%
\bibitem [{\citenamefont {Lee}(1977{\natexlab{a}})}]{lee19771}%
  \BibitemOpen
  \bibfield  {author} {\bibinfo {author} {\bibfnamefont {C.~T.}\ \bibnamefont
  {Lee}},\ }\bibfield  {title} {\bibinfo {title} {Exact solution of the
  superradiance master equation. {I}. {C}omplete initial excitation},\ }\href
  {https://doi.org/10.1103/PhysRevA.15.2019} {\bibfield  {journal} {\bibinfo
  {journal} {Phys. Rev. A}\ }\textbf {\bibinfo {volume} {15}},\ \bibinfo
  {pages} {2019} (\bibinfo {year} {1977}{\natexlab{a}})}\BibitemShut {NoStop}%
\bibitem [{\citenamefont {Lee}(1977{\natexlab{b}})}]{lee19772}%
  \BibitemOpen
  \bibfield  {author} {\bibinfo {author} {\bibfnamefont {C.~T.}\ \bibnamefont
  {Lee}},\ }\bibfield  {title} {\bibinfo {title} {Exact solution of the
  superradiance master equation. {II}. {A}rbitrary initial excitation},\ }\href
  {https://doi.org/10.1103/PhysRevA.16.301} {\bibfield  {journal} {\bibinfo
  {journal} {Phys. Rev. A}\ }\textbf {\bibinfo {volume} {16}},\ \bibinfo
  {pages} {301} (\bibinfo {year} {1977}{\natexlab{b}})}\BibitemShut {NoStop}%
\bibitem [{\citenamefont {Holzinger}\ and\ \citenamefont
  {Genes}(2025)}]{holzinger20252}%
  \BibitemOpen
  \bibfield  {author} {\bibinfo {author} {\bibfnamefont {R.}~\bibnamefont
  {Holzinger}}\ and\ \bibinfo {author} {\bibfnamefont {C.}~\bibnamefont
  {Genes}},\ }\bibfield  {title} {\bibinfo {title} {A compact analytical
  solution of the {Dicke} superradiance master equation via residue calculus},\
  }\href {https://doi.org/10.1515/zna-2025-0179} {\bibfield  {journal}
  {\bibinfo  {journal} {Zeitschrift für Naturforschung A}\ }\textbf {\bibinfo
  {volume} {80}},\ \bibinfo {pages} {673} (\bibinfo {year} {2025})}\BibitemShut
  {NoStop}%
\bibitem [{\citenamefont {Holzinger}\ \emph
  {et~al.}(2026{\natexlab{a}})\citenamefont {Holzinger}, \citenamefont
  {Bassler}, \citenamefont {Lyne}, \citenamefont {Jimenez}, \citenamefont
  {Gohsrich},\ and\ \citenamefont {Genes}}]{holzinger20251}%
  \BibitemOpen
  \bibfield  {author} {\bibinfo {author} {\bibfnamefont {R.}~\bibnamefont
  {Holzinger}}, \bibinfo {author} {\bibfnamefont {N.~S.}\ \bibnamefont
  {Bassler}}, \bibinfo {author} {\bibfnamefont {J.}~\bibnamefont {Lyne}},
  \bibinfo {author} {\bibfnamefont {F.~G.}\ \bibnamefont {Jimenez}}, \bibinfo
  {author} {\bibfnamefont {J.~T.}\ \bibnamefont {Gohsrich}},\ and\ \bibinfo
  {author} {\bibfnamefont {C.}~\bibnamefont {Genes}},\ }\bibfield  {title}
  {\bibinfo {title} {Solving {Dicke} superradiance analytically: A compendium
  of methods},\ }\href {https://doi.org/10.1103/9p44-kwfl} {\bibfield
  {journal} {\bibinfo  {journal} {Phys. Rev. A}\ }\textbf {\bibinfo {volume}
  {113}},\ \bibinfo {pages} {043718} (\bibinfo {year}
  {2026}{\natexlab{a}})}\BibitemShut {NoStop}%
\bibitem [{\citenamefont {Holzinger}\ \emph
  {et~al.}(2026{\natexlab{b}})\citenamefont {Holzinger}, \citenamefont
  {Bassler}, \citenamefont {Lyne}, \citenamefont {Yelin},\ and\ \citenamefont
  {Genes}}]{holzinger2026symbolic}%
  \BibitemOpen
  \bibfield  {author} {\bibinfo {author} {\bibfnamefont {R.}~\bibnamefont
  {Holzinger}}, \bibinfo {author} {\bibfnamefont {N.~S.}\ \bibnamefont
  {Bassler}}, \bibinfo {author} {\bibfnamefont {J.}~\bibnamefont {Lyne}},
  \bibinfo {author} {\bibfnamefont {S.~F.}\ \bibnamefont {Yelin}},\ and\
  \bibinfo {author} {\bibfnamefont {C.}~\bibnamefont {Genes}},\ }\bibfield
  {title} {\bibinfo {title} {Symbolic quantum-trajectory method for
  multichannel {Dicke} superradiance},\ }\href
  {https://doi.org/10.1103/5qcc-wfv6} {\bibfield  {journal} {\bibinfo
  {journal} {Phys. Rev. Lett.}\ }\textbf {\bibinfo {volume} {137}},\ \bibinfo
  {pages} {193602} (\bibinfo {year} {2026}{\natexlab{b}})}\BibitemShut
  {NoStop}%
\bibitem [{\citenamefont {Schwinger}(1952)}]{schwinger1952angular}%
  \BibitemOpen
  \bibfield  {author} {\bibinfo {author} {\bibfnamefont {J.}~\bibnamefont
  {Schwinger}},\ }\href {https://www.osti.gov/biblio/4389568} {\emph {\bibinfo
  {title} {On Angular Momentum}}},\ \bibinfo {type} {Tech. Rep.}\ \bibinfo
  {number} {NYO-3071}\ (\bibinfo  {institution} {Nuclear Development
  Associates, Inc.},\ \bibinfo {year} {1952})\BibitemShut {NoStop}%
\bibitem [{\citenamefont {Silva}\ and\ \citenamefont
  {Feist}(2022)}]{silva2022permutational}%
  \BibitemOpen
  \bibfield  {author} {\bibinfo {author} {\bibfnamefont {R.~E.~F.}\
  \bibnamefont {Silva}}\ and\ \bibinfo {author} {\bibfnamefont
  {J.}~\bibnamefont {Feist}},\ }\bibfield  {title} {\bibinfo {title}
  {Permutational symmetry for identical multilevel systems: A second-quantized
  approach},\ }\href {https://doi.org/10.1103/PhysRevA.105.043704} {\bibfield
  {journal} {\bibinfo  {journal} {Phys. Rev. A}\ }\textbf {\bibinfo {volume}
  {105}},\ \bibinfo {pages} {043704} (\bibinfo {year} {2022})}\BibitemShut
  {NoStop}%
\bibitem [{\citenamefont {Lyne}\ \emph {et~al.}(2026)\citenamefont {Lyne},
  \citenamefont {Schmidt}, \citenamefont {Genes},\ and\ \citenamefont
  {Bassler}}]{lyne2026}%
  \BibitemOpen
  \bibfield  {author} {\bibinfo {author} {\bibfnamefont {J.}~\bibnamefont
  {Lyne}}, \bibinfo {author} {\bibfnamefont {K.~P.}\ \bibnamefont {Schmidt}},
  \bibinfo {author} {\bibfnamefont {C.}~\bibnamefont {Genes}},\ and\ \bibinfo
  {author} {\bibfnamefont {N.~S.}\ \bibnamefont {Bassler}},\ }\href
  {https://arxiv.org/abs/2603.00778} {\bibinfo {title} {{Dicke} superradiance
  in degenerate quantum matter: Interplay of exchange statistics and spatial
  confinement}} (\bibinfo {year} {2026}),\ \Eprint
  {https://arxiv.org/abs/2603.00778} {arXiv:2603.00778 [cond-mat.quant-gas]}
  \BibitemShut {NoStop}%
\bibitem [{\citenamefont {Olivares}\ \emph {et~al.}(2007)\citenamefont
  {Olivares}, \citenamefont {Casagrande}, \citenamefont {Lulli},\ and\
  \citenamefont {Paris}}]{olivares2007}%
  \BibitemOpen
  \bibfield  {author} {\bibinfo {author} {\bibfnamefont {S.}~\bibnamefont
  {Olivares}}, \bibinfo {author} {\bibfnamefont {F.}~\bibnamefont
  {Casagrande}}, \bibinfo {author} {\bibfnamefont {A.}~\bibnamefont {Lulli}},\
  and\ \bibinfo {author} {\bibfnamefont {M.~G.~A.}\ \bibnamefont {Paris}},\
  }\bibfield  {title} {\bibinfo {title} {Reconstruction of the photon
  distribution in a micromaser},\ }\href
  {https://doi.org/10.1209/0295-5075/80/64002} {\bibfield  {journal} {\bibinfo
  {journal} {Europhysics Letters}\ }\textbf {\bibinfo {volume} {80}},\ \bibinfo
  {pages} {64002} (\bibinfo {year} {2007})}\BibitemShut {NoStop}%
\bibitem [{\citenamefont {Zambra}\ and\ \citenamefont
  {Paris}(2006)}]{zambra2006}%
  \BibitemOpen
  \bibfield  {author} {\bibinfo {author} {\bibfnamefont {G.}~\bibnamefont
  {Zambra}}\ and\ \bibinfo {author} {\bibfnamefont {M.~G.~A.}\ \bibnamefont
  {Paris}},\ }\bibfield  {title} {\bibinfo {title} {Reconstruction of
  photon-number distribution using low-performance photon counters},\ }\href
  {https://doi.org/10.1103/PhysRevA.74.063830} {\bibfield  {journal} {\bibinfo
  {journal} {Phys. Rev. A}\ }\textbf {\bibinfo {volume} {74}},\ \bibinfo
  {pages} {063830} (\bibinfo {year} {2006})}\BibitemShut {NoStop}%
\bibitem [{\citenamefont {Weyland}\ \emph {et~al.}(2025)\citenamefont
  {Weyland}, \citenamefont {Sanchez}, \citenamefont {Ruksasakchai},\ and\
  \citenamefont {Andersen}}]{weyland2025}%
  \BibitemOpen
  \bibfield  {author} {\bibinfo {author} {\bibfnamefont {M.}~\bibnamefont
  {Weyland}}, \bibinfo {author} {\bibfnamefont {L.}~\bibnamefont {Sanchez}},
  \bibinfo {author} {\bibfnamefont {P.}~\bibnamefont {Ruksasakchai}},\ and\
  \bibinfo {author} {\bibfnamefont {M.~F.}\ \bibnamefont {Andersen}},\
  }\bibfield  {title} {\bibinfo {title} {Applications of maximum likelihood
  estimations for analyzing photon counts in few-atom experiments},\ }\href
  {https://doi.org/10.1103/l616-hrf7} {\bibfield  {journal} {\bibinfo
  {journal} {Phys. Rev. A}\ }\textbf {\bibinfo {volume} {112}},\ \bibinfo
  {pages} {012817} (\bibinfo {year} {2025})}\BibitemShut {NoStop}%
\bibitem [{\citenamefont {Silberfarb}\ \emph {et~al.}(2005)\citenamefont
  {Silberfarb}, \citenamefont {Jessen},\ and\ \citenamefont
  {Deutsch}}]{silberfarb2005}%
  \BibitemOpen
  \bibfield  {author} {\bibinfo {author} {\bibfnamefont {A.}~\bibnamefont
  {Silberfarb}}, \bibinfo {author} {\bibfnamefont {P.~S.}\ \bibnamefont
  {Jessen}},\ and\ \bibinfo {author} {\bibfnamefont {I.~H.}\ \bibnamefont
  {Deutsch}},\ }\bibfield  {title} {\bibinfo {title} {Quantum state
  reconstruction via continuous measurement},\ }\href
  {https://doi.org/10.1103/PhysRevLett.95.030402} {\bibfield  {journal}
  {\bibinfo  {journal} {Phys. Rev. Lett.}\ }\textbf {\bibinfo {volume} {95}},\
  \bibinfo {pages} {030402} (\bibinfo {year} {2005})}\BibitemShut {NoStop}%
\bibitem [{\citenamefont {Rall}\ and\ \citenamefont {Wolf}(2026)}]{rall2026}%
  \BibitemOpen
  \bibfield  {author} {\bibinfo {author} {\bibfnamefont {H.}~\bibnamefont
  {Rall}}\ and\ \bibinfo {author} {\bibfnamefont {M.~M.}\ \bibnamefont
  {Wolf}},\ }\bibfield  {title} {\bibinfo {title} {Quantum tomography from the
  evolution of a single expectation},\ }\href
  {https://doi.org/10.1007/s00023-025-01617-4} {\bibfield  {journal} {\bibinfo
  {journal} {Annales Henri Poincar{\'e}}\ }\textbf {\bibinfo {volume} {27}},\
  \bibinfo {pages} {2939} (\bibinfo {year} {2026})}\BibitemShut {NoStop}%
\bibitem [{\citenamefont {Peruzzo}\ \emph {et~al.}(2026)\citenamefont
  {Peruzzo}, \citenamefont {Grigoletto},\ and\ \citenamefont
  {Ticozzi}}]{peruzzo2026}%
  \BibitemOpen
  \bibfield  {author} {\bibinfo {author} {\bibfnamefont {M.}~\bibnamefont
  {Peruzzo}}, \bibinfo {author} {\bibfnamefont {T.}~\bibnamefont
  {Grigoletto}},\ and\ \bibinfo {author} {\bibfnamefont {F.}~\bibnamefont
  {Ticozzi}},\ }\bibfield  {title} {\bibinfo {title} {Reconstructing quantum
  states and expectations via dynamical tomography},\ }\href
  {https://doi.org/10.1103/x88t-j8k9} {\bibfield  {journal} {\bibinfo
  {journal} {Phys. Rev. A}\ }\textbf {\bibinfo {volume} {113}},\ \bibinfo
  {pages} {032435} (\bibinfo {year} {2026})}\BibitemShut {NoStop}%
\bibitem [{\citenamefont {Mlynek}\ \emph {et~al.}(2014)\citenamefont {Mlynek},
  \citenamefont {Abdumalikov}, \citenamefont {Eichler},\ and\ \citenamefont
  {Wallraff}}]{mlynek2014}%
  \BibitemOpen
  \bibfield  {author} {\bibinfo {author} {\bibfnamefont {J.~A.}\ \bibnamefont
  {Mlynek}}, \bibinfo {author} {\bibfnamefont {A.~A.}\ \bibnamefont
  {Abdumalikov}, \bibfnamefont {Jr.}}, \bibinfo {author} {\bibfnamefont
  {C.}~\bibnamefont {Eichler}},\ and\ \bibinfo {author} {\bibfnamefont
  {A.}~\bibnamefont {Wallraff}},\ }\bibfield  {title} {\bibinfo {title}
  {Observation of {Dicke} superradiance for two artificial atoms in a cavity
  with high decay rate},\ }\href {https://doi.org/10.1038/ncomms6186}
  {\bibfield  {journal} {\bibinfo  {journal} {Nat. Commun.}\ }\textbf {\bibinfo
  {volume} {5}},\ \bibinfo {pages} {5186} (\bibinfo {year} {2014})}\BibitemShut
  {NoStop}%
\bibitem [{\citenamefont {Hansen}(2010)}]{hansen2010}%
  \BibitemOpen
  \bibfield  {author} {\bibinfo {author} {\bibfnamefont {P.~C.}\ \bibnamefont
  {Hansen}},\ }\href {https://doi.org/10.1137/1.9780898718836} {\emph {\bibinfo
  {title} {Discrete Inverse Problems: Insight and Algorithms}}}\ (\bibinfo
  {publisher} {Society for Industrial and Applied Mathematics},\ \bibinfo
  {address} {Philadelphia, PA},\ \bibinfo {year} {2010})\BibitemShut {NoStop}%
\bibitem [{\citenamefont {Chase}\ and\ \citenamefont
  {Geremia}(2008)}]{chase2008}%
  \BibitemOpen
  \bibfield  {author} {\bibinfo {author} {\bibfnamefont {B.~A.}\ \bibnamefont
  {Chase}}\ and\ \bibinfo {author} {\bibfnamefont {J.~M.}\ \bibnamefont
  {Geremia}},\ }\bibfield  {title} {\bibinfo {title} {Collective processes of
  an ensemble of spin-$\frac{1}{2}$ particles},\ }\href
  {https://doi.org/10.1103/PhysRevA.78.052101} {\bibfield  {journal} {\bibinfo
  {journal} {Phys. Rev. A}\ }\textbf {\bibinfo {volume} {78}},\ \bibinfo
  {pages} {052101} (\bibinfo {year} {2008})}\BibitemShut {NoStop}%
\bibitem [{\citenamefont {Baragiola}\ \emph {et~al.}(2010)\citenamefont
  {Baragiola}, \citenamefont {Chase},\ and\ \citenamefont
  {Geremia}}]{baragiola2010}%
  \BibitemOpen
  \bibfield  {author} {\bibinfo {author} {\bibfnamefont {B.~Q.}\ \bibnamefont
  {Baragiola}}, \bibinfo {author} {\bibfnamefont {B.~A.}\ \bibnamefont
  {Chase}},\ and\ \bibinfo {author} {\bibfnamefont {J.~M.}\ \bibnamefont
  {Geremia}},\ }\bibfield  {title} {\bibinfo {title} {Collective uncertainty in
  partially polarized and partially decohered spin-$\frac{1}{2}$ systems},\
  }\href {https://doi.org/10.1103/PhysRevA.81.032104} {\bibfield  {journal}
  {\bibinfo  {journal} {Phys. Rev. A}\ }\textbf {\bibinfo {volume} {81}},\
  \bibinfo {pages} {032104} (\bibinfo {year} {2010})}\BibitemShut {NoStop}%
\bibitem [{\citenamefont {Xu}\ \emph {et~al.}(2013)\citenamefont {Xu},
  \citenamefont {Tieri},\ and\ \citenamefont {Holland}}]{xu2013}%
  \BibitemOpen
  \bibfield  {author} {\bibinfo {author} {\bibfnamefont {M.}~\bibnamefont
  {Xu}}, \bibinfo {author} {\bibfnamefont {D.~A.}\ \bibnamefont {Tieri}},\ and\
  \bibinfo {author} {\bibfnamefont {M.~J.}\ \bibnamefont {Holland}},\
  }\bibfield  {title} {\bibinfo {title} {Simulating open quantum systems by
  applying {SU(4)} to quantum master equations},\ }\href
  {https://doi.org/10.1103/PhysRevA.87.062101} {\bibfield  {journal} {\bibinfo
  {journal} {Phys. Rev. A}\ }\textbf {\bibinfo {volume} {87}},\ \bibinfo
  {pages} {062101} (\bibinfo {year} {2013})}\BibitemShut {NoStop}%
\bibitem [{\citenamefont {Shammah}\ \emph {et~al.}(2018)\citenamefont
  {Shammah}, \citenamefont {Ahmed}, \citenamefont {Lambert}, \citenamefont
  {De~Liberato},\ and\ \citenamefont {Nori}}]{shammah2018}%
  \BibitemOpen
  \bibfield  {author} {\bibinfo {author} {\bibfnamefont {N.}~\bibnamefont
  {Shammah}}, \bibinfo {author} {\bibfnamefont {S.}~\bibnamefont {Ahmed}},
  \bibinfo {author} {\bibfnamefont {N.}~\bibnamefont {Lambert}}, \bibinfo
  {author} {\bibfnamefont {S.}~\bibnamefont {De~Liberato}},\ and\ \bibinfo
  {author} {\bibfnamefont {F.}~\bibnamefont {Nori}},\ }\bibfield  {title}
  {\bibinfo {title} {Open quantum systems with local and collective incoherent
  processes: Efficient numerical simulations using permutational invariance},\
  }\href {https://doi.org/10.1103/PhysRevA.98.063815} {\bibfield  {journal}
  {\bibinfo  {journal} {Phys. Rev. A}\ }\textbf {\bibinfo {volume} {98}},\
  \bibinfo {pages} {063815} (\bibinfo {year} {2018})}\BibitemShut {NoStop}%
\bibitem [{\citenamefont {Malz}\ \emph {et~al.}(2022)\citenamefont {Malz},
  \citenamefont {Trivedi},\ and\ \citenamefont {Cirac}}]{malz2022}%
  \BibitemOpen
  \bibfield  {author} {\bibinfo {author} {\bibfnamefont {D.}~\bibnamefont
  {Malz}}, \bibinfo {author} {\bibfnamefont {R.}~\bibnamefont {Trivedi}},\ and\
  \bibinfo {author} {\bibfnamefont {J.~I.}\ \bibnamefont {Cirac}},\ }\bibfield
  {title} {\bibinfo {title} {Large-{$N$} limit of {Dicke} superradiance},\
  }\href {https://doi.org/10.1103/PhysRevA.106.013716} {\bibfield  {journal}
  {\bibinfo  {journal} {Phys. Rev. A}\ }\textbf {\bibinfo {volume} {106}},\
  \bibinfo {pages} {013716} (\bibinfo {year} {2022})}\BibitemShut {NoStop}%
\bibitem [{\citenamefont {Gillespie}(1976)}]{gillespie1976general}%
  \BibitemOpen
  \bibfield  {author} {\bibinfo {author} {\bibfnamefont {D.~T.}\ \bibnamefont
  {Gillespie}},\ }\bibfield  {title} {\bibinfo {title} {A general method for
  numerically simulating the stochastic time evolution of coupled chemical
  reactions},\ }\href {https://doi.org/10.1016/0021-9991(76)90041-3} {\bibfield
   {journal} {\bibinfo  {journal} {J. Comput. Phys.}\ }\textbf {\bibinfo
  {volume} {22}},\ \bibinfo {pages} {403} (\bibinfo {year} {1976})}\BibitemShut
  {NoStop}%
\bibitem [{\citenamefont {Niehof}\ \emph {et~al.}(2026)\citenamefont {Niehof},
  \citenamefont {Ba{\ss}ler},\ and\ \citenamefont
  {Genes}}]{dicke_superradiance_reproducibility}%
  \BibitemOpen
  \bibfield  {author} {\bibinfo {author} {\bibfnamefont {C.}~\bibnamefont
  {Niehof}}, \bibinfo {author} {\bibfnamefont {N.~S.}\ \bibnamefont
  {Ba{\ss}ler}},\ and\ \bibinfo {author} {\bibfnamefont {C.}~\bibnamefont
  {Genes}},\ }\href
  {https://github.com/Nico0oblr/Code-for-Number-fluctuations-and-particle-loss-in-Dicke-superradiance-}
  {\bibinfo {title} {Code for number fluctuations and particle loss in {Dicke}
  superradiance}},\ \bibinfo {howpublished} {GitHub repository} (\bibinfo
  {year} {2026}),\ \bibinfo {note}
  {\url{https://github.com/Nico0oblr/Code-for-Number-fluctuations-and-particle-loss-in-Dicke-superradiance-}}\BibitemShut
  {NoStop}%
\end{thebibliography}%

\clearpage
\onecolumngrid

\appendix

\section{Higher moments}\label{app:moments}
Throughout this appendix, each occupied particle-number sector is initially fully excited. All relations expressing intensity derivatives solely in terms of particle-number moments assume this preparation. For a general initial density matrix, the initial intensity is $I(0)=\Gamma\langle\mop(\Nop-\mop+1)\rangle_0$. Under the fully excited preparation, Eq.~\eqref{eq:number_averaged_intensity} reduces to
\begin{equation}
    I(0)
    =\Gamma\braket{\hat S^\dagger \hat S}_0
    =\Gamma\Tr[\Nop\rho_0]
\end{equation}
The density matrix fulfills the equation of motion $\dot\rho=\mathcal L\rho$ with the Lindbladian $\mathcal L=\Gamma\mathcal D[\hat a_{\mathrm{g}}^\dagger\hat a_{\mathrm{e}}^{\phantom\dagger}]$.
Since the intensity operator $\hat S^\dagger\hat S$ does not depend explicitly on the time $t$, the initial derivative is given by
\begin{equation}
    \partial_tI(0)=\Gamma\Tr[\hat S^\dagger \hat S\mathcal L\rho_0].
\end{equation}
Using this equation recursively, the $k$-th derivative of intensity at $t=0$ is given by
\begin{equation}
    I^{(k)}(0)=\Gamma\Tr[\hat S^\dagger \hat S\mathcal L^k\rho_0].
\end{equation}
These derivatives can be computed analytically, giving the relations in Table~\ref{tab:HigherMoments}, which connect the $k$-th derivative to particle-number moments up to order $k+1$.
Furthermore, these relations can be inverted to obtain the $k$-th moment from $I(0)$ and derivatives up to order $k-1$.
The expressions for the first eight moments are given in Table~\ref{tab:HigherMomentsFromDerivatives}; all coefficients in these inverse relations are positive integers.

\begin{table}[ht]
    \centering
    \setlength{\tabcolsep}{1pt}
    \begin{tabular}{rrrrrrrrrrr}
        $I^{(0)}(0)/\Gamma^{\phantom 1}=$
        & $\braket{\Nop}$ \\[2pt]

        $I^{(1)}(0)/\Gamma^2=$
        & $-2\braket{\Nop}$
        & $+\braket{\Nop^2}$ \\[2pt]

        $I^{(2)}(0)/\Gamma^3=$
        & $8\braket{\Nop}$
        & $-8\braket{\Nop^2}$
        & $+\braket{\Nop^3}$ \\[2pt]

        $I^{(3)}(0)/\Gamma^4=$
        & $-56\braket{\Nop}$
        & $+76\braket{\Nop^2}$
        & $-22\braket{\Nop^3}$
        & $+\braket{\Nop^4}$ \\[2pt]

        $I^{(4)}(0)/\Gamma^5=$
        & $608\braket{\Nop}$
        & $-984\braket{\Nop^2}$
        & $+428\braket{\Nop^3}$
        & $-52\braket{\Nop^4}$
        & $+\braket{\Nop^5}$ \\[2pt]

        $I^{(5)}(0)/\Gamma^6=$
        & $-9\,440\braket{\Nop}$
        & $+17\,216\braket{\Nop^2}$
        & $-9\,560\braket{\Nop^3}$
        & $+1\,896\braket{\Nop^4}$
        & $-114\braket{\Nop^5}$
        & $+\braket{\Nop^6}$ \\[2pt]

        $I^{(6)}(0)/\Gamma^7=$
        & $198\,272\braket{\Nop}$
        & $-395\,200\braket{\Nop^2}$
        & $+259\,168\braket{\Nop^3}$
        & $-69\,344\braket{\Nop^4}$
        & $+7\,344\braket{\Nop^5}$
        & $-240\braket{\Nop^6}$
        & $+\braket{\Nop^7}$ \\[2pt]

        $I^{(7)}(0)/\Gamma^8=$
        & $-5\,410\,688\braket{\Nop}$
        & $+11\,566\,656\braket{\Nop^2}$
        & $-8\,576\,480\braket{\Nop^3}$
        & $+2\,817\,456\braket{\Nop^4}$
        & $-422\,648\braket{\Nop^5}$
        & $+26\,196\braket{\Nop^6}$
        & $-494\braket{\Nop^7}$
        & $+\braket{\Nop^8}$ \\[2pt]

    \end{tabular}
    \caption{Initial intensity derivatives expressed in terms of emitter number statistics moments.}
    \label{tab:HigherMoments}
\end{table}

\newcommand{\I}[2]{\iota_{#1}}
\begin{table}[ht]
    \centering
    \begin{tabular}{rrrrrrrrrrrrr}
        $\braket{\Nop}=$
        & $\I01$\\
        $\braket{\Nop^2}=$
        & $2\I01$
        & $+\I12$\\
        $\braket{\Nop^3}=$
        & $8\I01$
        & $+8\I12$
        & $+\I23$\\
        $\braket{\Nop^4}=$
        & $80\I01$
        & $+100\I12$
        & $+22\I23$
        & $+\I34$\\
        $\braket{\Nop^5}=$
        & $2\,096\I01$
        & $+2\,760\I12$
        & $+716\I23$
        & $+52\I34$
        & $+\I45$\\
        $\braket{\Nop^6}=$
        & $138\,752\I01$
        & $+184\,304\I12$
        & $+49\,472\I23$
        & $+4\,032\I34$
        & $+114\I45$
        & $+\I56$\\
        $\braket{\Nop^7}=$
        & $21\,973\,760\I01$
        & $+29\,219\,776\I12$
        & $+7\,881\,376\I23$
        & $+655\,136\I34$
        & $+20\,016\I45$
        & $+240\I56$
        & $+\I67$\\
        $\braket{\Nop^8}=$
        & $7\,931\,652\,992\I01$
        & $+10\,548\,349\,824\I12$
        & $+2\,846\,639\,648\I23$
        & $+237\,175\,152\I34$
        & $+7\,324\,208\I45$
        & $+92\,364\I56$
        & $+494\I67$
        & $+\I78$\\
    \end{tabular}
    \caption{Moments of the initial number distribution expressed through initial intensity derivatives, $\iota_k\coloneqq I^{(k)}(0)/\Gamma^{k+1}$.}
    \label{tab:HigherMomentsFromDerivatives}
\end{table}

\section{Particle distribution details}\label{app:dist}

A coherent state of excited atoms is given by
\begin{equation}
    \ket{\alpha}_{\mathrm{e}}=\me^{-|\alpha|^2/2}\sum_{N=0}^{\infty}\frac{\alpha^N}{\sqrt{N!}}\ket{N}_{\mathrm{e}}.
\end{equation}
As examples with broader and strongly non-Poissonian number statistics, we consider the cat state
\begin{equation}
    \ket{\psi_{\mathrm{cat}}}
    =\frac{\ket{\alpha}_{\mathrm{e}}+\ket{-\alpha}_{\mathrm{e}}}
    {\sqrt{2\left(1+\me^{-2|\alpha|^2}\right)}}\otimes\ket{0}_{\mathrm{g}},
\end{equation}
and a squeezed vacuum state with real squeezing parameter $s\geq0$,
\begin{equation}
    \ket{s}
    =\frac{1}{\sqrt{\cosh s}}
    \sum_{k=0}^{\infty}
    \frac{(-\tanh s)^k\sqrt{(2k)!}}{2^k k!}
    \ket{2k}_{\mathrm{e}}\otimes\ket{0}_{\mathrm{g}}.
\end{equation}
Both states have support only on even particle numbers. We furthermore consider a thermal state,
\begin{equation}
    \rho_{\mathrm{th}}
    =(1-\me^{-\delta})
    \sum_{N=0}^{\infty}\me^{-\delta N}
    \outerproduct{N}{N}_{\mathrm{e}}\otimes\outerproduct{0}{0}_{\mathrm{g}},
\end{equation}
where $\delta=\hbar\omega_0/(k_{\mathrm{B}} T)$.

\section{Conditioning of neighboring intensity curves}\label{app:intensity_cond}

The neighboring fixed-$N$ intensity curves $I_N(t)$ and $I_{N+1}(t)$ become increasingly similar as $N$ grows, worsening the conditioning of the linear system. This can be seen by considering the overlap
\begin{equation}
    \mathcal O_N=\frac{\int_0^\infty I_N(t)I_{N+1}(t)\dd t}{\sqrt{\int_0^\infty I_N^2(t)\dd t\int_0^\infty I_{N+1}^2(t)\dd t}},
\end{equation}
for which the mean-field $\operatorname{sech}^2$ pulse from App.~\ref{app:dicke_mean_field} gives the large-$N$ estimate
\begin{equation}\label{eq:overlap}
    \mathcal O^{\mathrm{MF}}_N=1-\frac{1}{2N^2}\left[\frac{(\ln N-1)^2}{5}+\frac{\pi^2}{15}-\frac14\right]+O\left(\frac{1}{N^3}\right),
\end{equation}
indeed confirming that $\lim\limits_{N\to\infty}\mathcal O^{\mathrm{MF}}_N=1$. The coefficient of the $O(\ln^2N/N^2)$ correction is specific to this mean-field estimate: the delay-averaged pulse $F(z)$ in App.~\ref{app:stochastic_asymptotics} changes this coefficient but preserves the limit of unity.

\section{More details on the reconstruction of statistics}\label{app:rec}

\begin{figure}
    \centering
    \includegraphics[width=\linewidth]{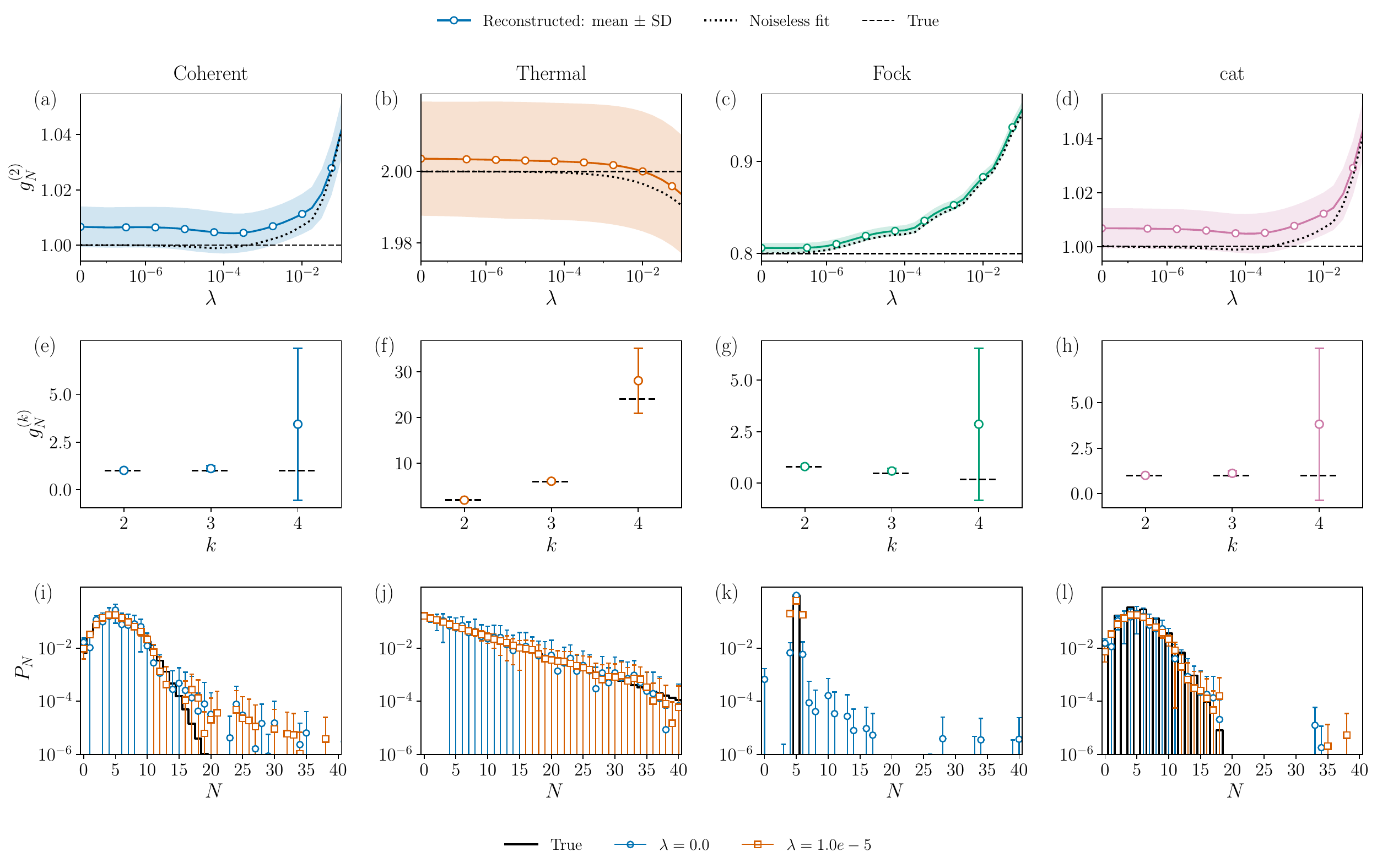}
    \caption{Sensitivity of particle-number reconstruction to smoothing and moment order. Columns show coherent, thermal, Fock, and cat states with mean occupation $\langle\hat N\rangle=5$. (a--d) Reconstructed $g_{\mathrm{N}}^{(2)}$ versus the second-difference penalty strength $\lambda$; dotted curves show reconstructions from noiseless ODE intensities. (e--h) Reconstructed normalized factorial moments $g_{\mathrm{N}}^{(k)}$ for $k=2,3,4$, using $\lambda=0$. Dashed lines and segments indicate the true moments. (i--l) True distributions (black steps) and reconstructed probabilities for $\lambda=0$ and $10^{-5}$. Reconstructed values are averaged over 30 independent runs, each containing $R=10^5$ preparations; shading and error bars indicate one sample standard deviation across runs. The same observations are reused at each $\lambda$. Simulations include independent photon loss with $\eta=0.9$ and an exponential detector response with $\Gamma\tau=0.05$, using $\Gamma T_{\mathrm{acq}}=20$, $\Gamma\Delta t=0.01$, and zero background. Reconstruction uses nonnegative, normalized least squares on $0\leq N\leq150$; only $N\leq40$ is displayed in the bottom row.}
\label{fig:rec_app}
\end{figure}

In Fig.~\ref{fig:rec_app} we show a more detailed study of the reconstruction of emitter-number moments and populations from the intensity curve, as defined in the main text. We investigate the impact of the smoothing parameter for four states: coherent, thermal, Fock, and cat, respectively, in the first row. For the distributions with broader population profiles, including the cat state with its alternating parity structure, weak smoothing has little impact on the reconstructed \(g_{\mathrm{N}}^{(2)}\). For the sharply localized Fock state, however, it introduces a systematic bias starting at around \(\lambda=10^{-6}\), as also indicated by the noiseless reconstructions. While smoothness regularization can reduce statistical fluctuations by suppressing poorly constrained features of the distribution, it simultaneously introduces prior assumptions about its structure. Consequently, without more specific knowledge of the initial distribution, the dependence on the smoothing parameter should be checked over several values.

In the second row, we reconstruct higher-order correlation functions defined through the factorial moments

\begin{equation}
    F_k=\left\langle(\Nop)_k\right\rangle=\sum_{N=k}^{\infty}\frac{N!}{(N-k)!}P_N,
    \qquad
    g_{\mathrm{N}}^{(k)}=\frac{F_k}{\langle\Nop\rangle^k}.
\end{equation}
Here, $(\Nop)_k=\Nop(\Nop-1)\cdots(\Nop-k+1)$ denotes the falling factorial.

The reconstruction works well for \(g_{\mathrm{N}}^{(2)}\), as established in the main text, and remains reasonably accurate for \(g_{\mathrm{N}}^{(3)}\). However, the accuracy deteriorates dramatically for \(g_{\mathrm{N}}^{(4)}\). The reason becomes apparent in the bottom row, where we show the population reconstructions for two smoothing values at a larger mean occupation, \(\langle\hat N\rangle=5\), compared with the main text. Indeed, the reconstructions tend to contain spurious small contributions at large \(N\), which are poorly constrained by the intensity measurements at the available statistical precision. Although these contributions have little impact on the lower moments, they become increasingly important for higher-order correlations due to the factorial weights, which scale as \(N^k\) for large \(N\). Thus, even when the population reconstruction is inaccurate, low-order correlations can remain comparatively robust.

One may introduce additional heuristics, for instance by assuming a sufficiently rapidly decaying distribution and setting reconstructed contributions below some noise floor to zero. This would suppress the spurious large-\(N\) populations and can be expected to improve higher-order moments for sufficiently localized distributions. However, we also see that such a procedure could worsen the reconstruction of broader distributions, such as the thermal state, where small populations at large \(N\) constitute a genuine contribution to the higher moments. More generally, the results demonstrate that additional prior knowledge of the distribution can substantially improve the reconstruction by restricting or regularizing the optimization, but that the appropriate assumptions depend on the underlying number statistics.

\section{Derivation of the particle-loss operator: local loss}\label{app:loss_model}

We derive here the correct loss operator for a model of emitters to which local loss reservoirs are attached, as for instance emitters in an optical lattice in a cavity.  We define as underlying system a Hilbert space of three-level emitters with levels $\ket{e},\ket{g},\ket{a}$ where $\ket{a}$ is an auxiliary state that does not participate in the rest of the system dynamics. We thus define two Hilbert spaces
\begin{equation}
    \mathcal H_{\mathrm{e,g}}=\bigoplus_{N=0}^\infty \mathcal H_{\mathrm{e,g}}^{(N)},\qquad \mathcal H_{\mathrm{e,g,a}}=\bigoplus_{N=0}^\infty \mathcal H_{\mathrm{e,g,a}}^{(N)}
\end{equation}
where
\begin{equation}
        \mathcal H_{\mathrm{e,g}}^{(N)}=\text{span}\left\{\bigotimes_{q=1}^N \ket{\alpha_q}\,\middle\vert\,\alpha_q\in\{\mathrm{e},\mathrm{g}\}\right\},\quad\text{and}\quad
        \mathcal H_{\mathrm{e,g,a}}^{(N)}=\text{span}\left\{\bigotimes_{q=1}^N \ket{\alpha_q}\,\middle\vert\,\alpha_q\in\{\mathrm{e},\mathrm{g},\mathrm{a}\}\right\}.
\end{equation}

\begin{figure}[!ht]
    \centering
    \includegraphics[width=0.9\textwidth]{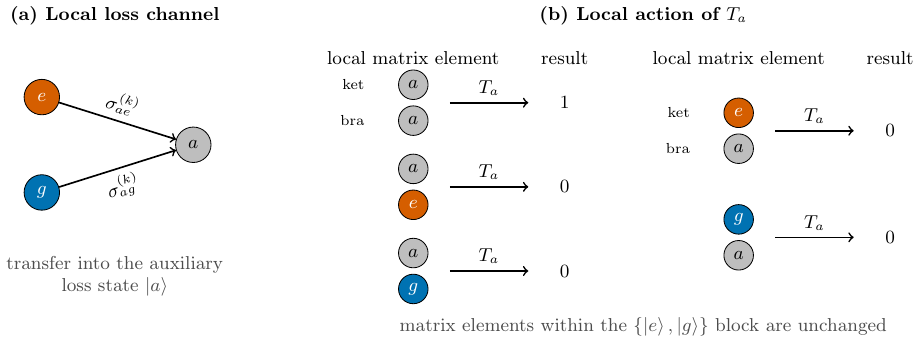}
\caption{Local particle loss and trace operation. (a) Excited- and ground-state emitters are transferred to the auxiliary loss state $\ket{a}$. (b) The trace $T_{\mathrm{a}}$ deletes matched auxiliary-state slots and removes matrix elements with mismatched auxiliary-state records.}
\label{fig:tikz1}
\end{figure}

The linear particle loss operators $\sigma^{(k)}_{\mathrm{ae}}$ and $\sigma^{(k)}_{\mathrm{ag}}$ are then defined as linear maps
\begin{equation}
    \sigma^{(k)}_{\mathrm{ae}}:\mathcal H_{\mathrm{e,g}}^{(N)}\to \mathcal H_{\mathrm{e,g,a}}^{(N)},\qquad \sigma^{(k)}_{\mathrm{ag}}:\mathcal H_{\mathrm{e,g}}^{(N)}\to \mathcal H_{\mathrm{e,g,a}}^{(N)}
\end{equation}
defined as
\begin{equation}
\begin{aligned}
    \sigma_{\mathrm{ae}}^{(k)}
    \ket{\alpha_1,\ldots,\alpha_N}
    &=
    \delta_{\alpha_k,\mathrm{e}}
    \ket{\alpha_1,\ldots,\alpha_{k-1},a,\alpha_{k+1},\ldots,\alpha_N},\\
    \sigma_{\mathrm{ag}}^{(k)}
    \ket{\alpha_1,\ldots,\alpha_N}
    &=
    \delta_{\alpha_k,\mathrm{g}}
    \ket{\alpha_1,\ldots,\alpha_{k-1},a,\alpha_{k+1},\ldots,\alpha_N},
\end{aligned}
\end{equation}

We also define a linear trace operation for density operators that removes emitters in the auxiliary state in an incoherent manner
\begin{equation}
    T_{\mathrm{a}}:\text{Hom}(\mathcal H_{\mathrm{e,g,a}}^{N'},\mathcal H_{\mathrm{e,g,a}}^{N})\to \bigoplus_{r=0}^{\min(N,N')}\text{Hom}(\mathcal H_{\mathrm{e,g}}^{N'-r}, \mathcal H_{\mathrm{e,g}}^{N-r})
\end{equation}
where $\text{Hom}(A,B)$ denotes the vector space of linear maps from $A$ to $B$.

If we define the set of auxiliary-state slots as $A(\alpha_1,\ldots,\alpha_N)=\{j\in\{1,\ldots,N\}\,\vert\, \alpha_j=\mathrm{a}\}$ and let $\bar A(\alpha_1,\ldots,\alpha_N)$ denote the sequence $(\alpha_j)_{\alpha_j\neq\mathrm{a}}$ of remaining internal states in their original order, then the trace operation is defined as
\begin{equation}
    T_{\mathrm{a}}(\ket{\alpha_1,\ldots,\alpha_N}\bra{\beta_1,\ldots,\beta_{N'}})=\begin{cases}
        \ket{\bar A(\alpha_1,\ldots,\alpha_N)}\bra{\bar A(\beta_1,\ldots,\beta_{N'})}\text{ for }A(\alpha_1,\ldots,\alpha_N)=A(\beta_1,\ldots,\beta_{N'})\\
        0\text{ else}
    \end{cases}
\end{equation}
which simply implements the local rules
\begin{equation}
    T_{\mathrm{a}}(\outerproduct{a}{e})\to 0, \quad T_{\mathrm{a}}(\outerproduct{a}{g})\to 0, \quad T_{\mathrm{a}}(\outerproduct{a}{a})\to 1,
\end{equation}
with the same rules for the Hermitian conjugates.

\begin{figure}[!ht]
    \includegraphics[width=0.7\textwidth]{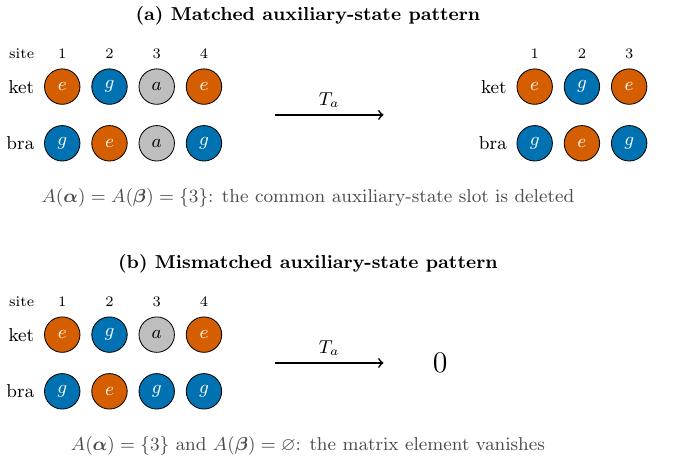}
\caption{Action of the trace $T_{\mathrm{a}}$ on many-body matrix elements. Matching auxiliary-state patterns on the ket and bra sides yield a reduced matrix element with the common auxiliary-state slot removed, while mismatched patterns vanish.}
\label{fig:tikz2}
\end{figure}

The trace is defined on general density matrices by linearity. The particle loss that mixes loss at equal rates over all particles in the system is then defined as
\begin{equation}
    J_{\mathrm{e}}[\rho]=T_{\mathrm{a}}\left[\sum_{k=1}^{\infty}\sigma_{\mathrm{ae}}^{(k)}\rho\left(\sigma_{\mathrm{ae}}^{(k)}\right)^\dagger\right],\qquad
    J_{\mathrm{g}}[\rho]=T_{\mathrm{a}}\left[\sum_{k=1}^{\infty}\sigma_{\mathrm{ag}}^{(k)}\rho\left(\sigma_{\mathrm{ag}}^{(k)}\right)^\dagger\right].
\end{equation}

We now determine the action of these operators on the Dicke states in \(\mathcal H_{\mathrm{e,g}}^{(N)}\), defined by
\begin{equation}
    \ket{m}_N=\frac{1}{\sqrt{\binom{N}{m}}}
    \sum_{\substack{\alpha_1,\ldots,\alpha_N\in\{\mathrm{e},\mathrm{g}\}\\
    \sum_{k=1}^N\delta_{\alpha_k,\mathrm{e}}=m}}
    \ket{\alpha_1,\ldots,\alpha_N},
\end{equation}
where \(m\) denotes the number of excited emitters. Note that for any site $j=1,\ldots,N$ one can decompose this state as
\begin{equation}\label{eq:Dicke_decomp}
    \ket{m}_{N}=\sqrt{\frac{N-m}{N}}\ket{m}_{N-1}\otimes\ket{g}_{j}+\sqrt{\frac{m}{N}}\ket{m-1}_{N-1}\otimes\ket{e}_{j}.
\end{equation}
up to ordering of the tensor factors.

\begin{figure}[!ht]
    \centering
    \includegraphics[width=0.9\textwidth]{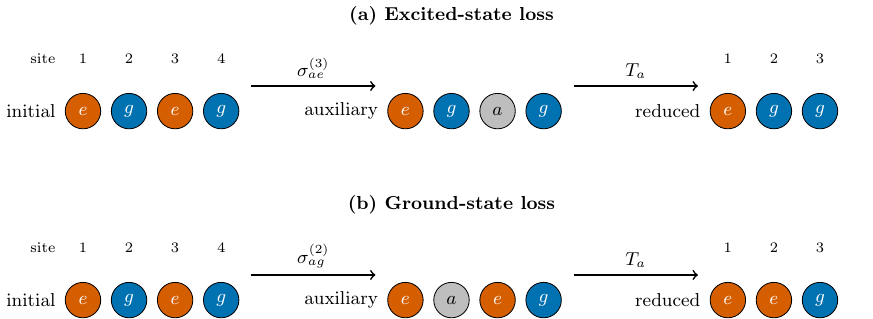}
\caption{Two-step construction of particle loss. The local operator $\sigma_{\mathrm{a}\alpha}^{(j)}$ first transfers the emitter at site $j$ to the loss state $\ket{a}$, after which $T_{\mathrm{a}}$ removes the auxiliary-state slot and maps the state to the corresponding $(N-1)$-particle sector.}
\label{fig:tikz3}
\end{figure}

Using this expression for the excited-state loss operator
\begin{equation}
\begin{aligned}
    J_{\mathrm{e}}\left[\ket{m}_N\bra{m'}_{N'}\right]
    &=
    T_{\mathrm{a}}\left[
    \sum_{j=1}^{\infty}
    \sigma_{\mathrm{ae}}^{(j)}
    \ket{m}_N\bra{m'}_{N'}
    \left(\sigma_{\mathrm{ae}}^{(j)}\right)^\dagger
    \right]\\
    &=
    \sqrt{\frac{m}{N}}\sqrt{\frac{m'}{N'}}
    \sum_{j=1}^{\min(N,N')}
    T_{\mathrm{a}}\left[
    \left(\ket{m-1}_{N-1}\otimes\ket{a}_j\right)
    \left(\bra{m'-1}_{N'-1}\otimes\bra{a}_j\right)
    \right]\\
    &=
    \min(N,N')\sqrt{\frac{mm'}{NN'}}
    \ket{m-1}_{N-1}\bra{m'-1}_{N'-1}.
\end{aligned}
\end{equation}

By a similar calculation, one finds
\begin{equation}
        J_{\mathrm{g}}\left[\ket{m}_N\bra{m'}_{N'}\right]=\min(N,N')\sqrt{\frac{(N-m)(N'-m')}{NN'}}\ket{m}_{N-1}\bra{m'}_{N'-1}.
\end{equation}

The remaining terms follow by noting that
\begin{equation}
    \sum_{j=1}^{\infty}\left(\sigma_{\mathrm{ae}}^{(j)}\right)^\dagger\sigma_{\mathrm{ae}}^{(j)}
    =\sum_{j=1}^{\infty}\ket{e}_j\bra{e}_j,
\end{equation}
which is the excitation-number operator $\mop$. Hence,
\begin{equation}
\begin{aligned}
    \sum_{j=1}^{\infty}\left(\sigma_{\mathrm{ae}}^{(j)}\right)^\dagger\sigma_{\mathrm{ae}}^{(j)}
    \ket{m}_N\bra{m'}_{N'}
    &=m\ket{m}_N\bra{m'}_{N'},\\
    \ket{m}_N\bra{m'}_{N'}
    \sum_{j=1}^{\infty}\left(\sigma_{\mathrm{ae}}^{(j)}\right)^\dagger\sigma_{\mathrm{ae}}^{(j)}
    &=m'\ket{m}_N\bra{m'}_{N'}.
\end{aligned}
\end{equation}
Similarly,
\begin{equation}
    \sum_{j=1}^{\infty}\left(\sigma_{\mathrm{ag}}^{(j)}\right)^\dagger\sigma_{\mathrm{ag}}^{(j)}
    =\sum_{j=1}^{\infty}\ket{g}_j\bra{g}_j,
\end{equation}
so that
\begin{equation}
\begin{aligned}
    \sum_{j=1}^{\infty}\left(\sigma_{\mathrm{ag}}^{(j)}\right)^\dagger\sigma_{\mathrm{ag}}^{(j)}
    \ket{m}_N\bra{m'}_{N'}
    &=(N-m)\ket{m}_N\bra{m'}_{N'},\\
    \ket{m}_N\bra{m'}_{N'}
    \sum_{j=1}^{\infty}\left(\sigma_{\mathrm{ag}}^{(j)}\right)^\dagger\sigma_{\mathrm{ag}}^{(j)}
    &=(N'-m')\ket{m}_N\bra{m'}_{N'}.
\end{aligned}
\end{equation}

So we see that for the full dissipator the naive replacement 
\begin{equation}
\begin{aligned}
    \sum_{q=1}^\infty  \mathcal D\left[\sigma_{\mathrm{ae}}^{(q)}\right]&\rightarrow \mathcal D\left[\hat a_{\mathrm{e}}\right]\\
    \sum_{q=1}^\infty  \mathcal D\left[\sigma_{\mathrm{ag}}^{(q)}\right]&\rightarrow \mathcal D\left[\hat a_{\mathrm{g}}\right]
\end{aligned}
\end{equation}
is valid for matrix elements $\ket{m}_{N}\bra{m'}_{N'}$ with $N=N'$. For coherences between different particle-number sectors, the local-loss model suppresses the recycling term.

\section{Mean-field solution}
\label{app:mean_field}

Here we use the hydrodynamic mean-field approximation to solve the equations of motion for Dicke superradiance, with and without atom loss, in the large-$N$ limit.

\subsection{Dicke mean-field equations}
\label{app:dicke_mean_field}

Keeping the notation from the main text, for Dicke superradiance of a system with $N$ emitters we restate the equations of motion for the Dicke populations
\begin{equation}
\frac{d p_m^{(N)}}{dt}=\Gamma h_{m+1}^{(N)}p_{m+1}^{(N)}-\Gamma h_m^{(N)}p_m^{(N)}.
\end{equation}

We then find the equation of motion for $\expval{\mop}$ as
\begin{equation}
\begin{aligned}
        \dv{\expval{\mop}}{t}&=\Gamma\sum_{m=0}^N m\left[h_{m+1}^{(N)}p^{(N)}_{m+1}-h_m^{(N)}p_m^{(N)}\right]\\
        &=-\Gamma\sum_{m=0}^N h_m^{(N)}p_m^{(N)}\\
        &=-\Gamma[(N+1)\expval{\mop}-\expval{\mop^2}]
\end{aligned}
\end{equation}

Now, we make the approximate substitution $\expval{\mop^2}\approx \expval{\mop}^2$ and arrive at the equation of motion
\begin{equation}
    -I_N(t)=\dv{\expval{\mop}}{t}\approx -\Gamma[(N+1)\expval{\mop}-\expval{\mop}^2]
\end{equation}
which is the usual mean-field result, but with $m$ representing the excitation number instead of the usual magnetic quantum number. The intensity is found via the derivative. The solution is known
\begin{equation}
\begin{aligned}
    \expval{\mop}(t)&=\frac{N+1}{2}\left[1-\tanh\left(\frac{\Gamma(N+1)}{2}(t-t_N)\right)\right],\\
    I_N(t)&=\frac{\Gamma(N+1)^2}{4}\operatorname{sech}^2\left(\frac{\Gamma(N+1)}{2}(t-t_N)\right).
\end{aligned}
\end{equation}
where $t_N=\frac{\ln N}{\Gamma(N+1)}\approx \frac{\ln N}{\Gamma N}$.

\subsection{Atom loss mean-field equations}
\label{app:loss_mean_field}

Including atom loss, the mean field follows from the same idea. Start with the equation of motion for the population distribution
\begin{equation}
\begin{aligned}
\frac{d}{dt}p_m^{(N)}(t)&=\Gamma h_{m+1}^{(N)}p_{m+1}^{(N)}(t)-\Gamma h_m^{(N)}p_m^{(N)}(t)\\
&+r(m+1)\,p_{m+1}^{(N+1)}(t)+r(N+1-m)\,p_m^{(N+1)}(t)-rN\,p_m^{(N)}(t).
\end{aligned}
\end{equation}

Then, we derive the equations of motion for $\expval{\mop}$ and $\expval{\Nop}$ from these rate equations. This yields
\begin{equation}
\begin{aligned}
    \dv{\expval{\mop}}{t}&=-\Gamma[\expval{(\Nop+1)\mop}-\expval{\mop^2}]-r\expval{\mop}\\
    \dv{\expval{\Nop}}{t}&=-r\expval{\Nop}.
\end{aligned}
\end{equation}

Making the mean-field approximation for the excitation number again, we obtain

\begin{equation}
\begin{aligned}
    \dv{\expval{\mop}}{t}&\approx-\Gamma[(\expval{\Nop}+1)\expval{\mop}-\expval{\mop}^2]-r\expval{\mop}\\
    \dv{\expval{\Nop}}{t}&=-r\expval{\Nop}.
\end{aligned}
\end{equation}

Writing $N_0=\expval{\Nop}_0$, we have $\expval{\Nop}(t)=N_0\me^{-rt}$ and hence $ \dv{\expval{\mop}}{t}\approx-\Gamma[(\expval{\Nop}_0\me^{-rt}+1)\expval{\mop}-\expval{\mop}^2]-r\expval{\mop}$.

We substitute $\expval{\mop}=\me^{-rt}u$ and $\tilde t=\frac{\Gamma}{r}(1-\me^{-rt})$ which yields
\begin{equation}
    \dv{u}{\tilde t}=u^2-\left[N_0+\frac{1}{1-\tilde t r/\Gamma}\right]u
\end{equation}
which is exactly integrable 
\begin{equation}
u(\tilde t)=\frac{u_0 \me^{-N_0\tilde t}\left(1-\frac{r\tilde t}{\Gamma}\right)^{\Gamma/r}}{1-u_0\int_0^{\tilde t} \me^{-N_0s}\left(1-\frac{rs}{\Gamma}\right)^{\Gamma/r}\dd{s}}.
\end{equation}

Back-substituting then yields
\begin{equation}
    \begin{aligned}
    \expval{\mop}(t)&=\frac{N_0 \me^{-\left(\frac{\Gamma N_0}{r}\left(1-\me^{-rt}\right)+(\Gamma+r)t\right)}}{1-\Gamma N_0\int_0^t \me^{-\left(\frac{\Gamma N_0}{r}\left(1-\me^{-rs}\right)+(\Gamma+r)s\right)}\dd s},\\
    \expval{\Nop}(t)&=N_0\me^{-rt},\\
    I(t)&=\Gamma\left[\expval{\Nop}(t)+1-\expval{\mop}(t)\right]\expval{\mop}(t).
    \end{aligned}
\end{equation}

\subsection{Scaling laws for large particle number}
\label{app:large_n_scaling}
In this subsection, $m(\bar t)=\expval{\mop}(t)$ and $N(\bar t)=\expval{\Nop}(t)$ denote scalar mean-field values at rescaled time $\bar t=rt$, while the bare symbol $N=N(0)$ denotes the initial particle number.
The fraction of excited particles
\begin{equation}
    p(\bar t)
    \coloneqq\frac{m(\bar t)}{N(\bar t)}
    =\frac{E(\bar t)\ee^{-q\bar t}}{1-a\int_0^{\bar t} E(u)\ee^{-(1+q)u}\dd u}
\end{equation}
is expressed in terms of
\begin{equation}
    E(\bar t)\coloneqq \exp[-a(1-\ee^{-\bar t})]
    \quad\text{and parameters}\quad
    a\coloneqq \ln N/g_{\mathrm{r}},\quad
    q\coloneqq a/N
\end{equation}
with additional rescaling $u=rs$ of the integration time $s$.
The scaling parameter $g_{\mathrm{r}}\coloneqq r\ln N/(\Gamma N)$ is kept constant throughout this calculation.
The identity
\begin{equation}
    E(\bar t)\ee^{-q\bar t}-1
    =\int_0^{\bar t}\frac{\dd}{\dd u}\bigl[E(u)\ee^{-qu}\bigr]\dd u
    =-a\int_0^{\bar t} E(u)\ee^{-(q+1)u}\dd u-q\int_0^{\bar t} E(u)\ee^{-qu}\dd u
\end{equation}
is used to rewrite the denominator in the expression for $p$.
It yields
\begin{equation}
    \frac1{p(\bar t)}
    =1+q\int_0^{\bar t}\frac{E(u)}{E(\bar t)}\frac{\ee^{-qu}}{\ee^{-q\bar t}}\dd u
    =1+q\exp[-a\ee^{-\bar t}+q\bar t]\int_0^{\bar t}\exp[a\ee^{-u}-qu]\dd u.
\end{equation}
In the limit of large particle numbers $N\rightarrow\infty$, the scaling behavior of the parameters $a\propto\ln N$ and $q\propto\ln N/N$ leads to a domination of the $a$-term in the asymptotic limit
\begin{equation}
    \frac1{p(\bar t)}
    \sim1+q\exp(-a\ee^{-\bar t})\int_0^{\bar t}\exp(a\ee^{-u})\dd u
\end{equation}
Moreover, the integrand becomes maximal at the origin $u=0$ and it is thus expanded around its maximum.
The resulting expression can be approximated by $\int_0^{\bar t}\exp(a\ee^{-u})\dd u\approx\int_0^\infty\ee^{a-au}\dd u=\ee^a/a$.
The final expression for the asymptotic limit of the fraction of excited particles becomes
\begin{equation}
    p(\bar t)
    \sim\frac1{1+\exp[a(1-\ee^{-\bar t})-\ln N]}.
\end{equation}
The maximal intensity thus scales as
\begin{equation}
    I_{\mathrm{max}}
    =\sup_{\bar t\in[0,\infty)}\Bigl\{\Gamma m(\bar t)\bigl[N(\bar t)-m(\bar t)+1\bigr]\Bigr\}
    \sim\Gamma N^2\sup_{\bar t\in[0,\infty)}\Bigl\{\ee^{-2\bar t}p(\bar t)\bigl[1-p(\bar t)\bigr]\Bigr\}
\end{equation}

\subsubsection{Dicke regime}
In the case of $0\leq g_{\mathrm{r}}<1$, a Dicke superradiance peak is seen in the intensity.
Thus, the fraction of excited atoms $p(0)\sim1$ starts initially at one and vanishes for large times $\lim_{\bar t\rightarrow\infty}p(\bar t)=0$.
The product $p(\bar t)\bigl[1-p(\bar t)\bigr]$ vanishes initially and for long times.
It is maximal for $p(\bar t)=1/2$, which occurs at $\bar t_{\mathrm{max}}=-\ln(1-g_{\mathrm{r}})\in[0,\infty)$.
At this time, the intensity
\begin{equation}
    I(\bar t_{\mathrm{max}})
    \sim\Gamma\frac{(1-g_{\mathrm{r}})^2}4N^2
    \sim I_{\mathrm{max}}
\end{equation}
scales quadratically with the particle number $N$ as for pure Dicke superradiance.

\subsubsection{Critical value}
At the critical scaling parameter $g_{\mathrm{r}}=1$, the intensity scales as
\begin{equation}
    I\sim
    \Gamma\frac{x^2}{4\cosh^2(x/2)}\,\frac{N^2}{\ln^2N},
\end{equation}
with $x(\bar t)=\ee^{-\bar t}\ln N$.
The intensity is maximal for $2=x_{\mathrm{max}}\tanh(x_{\mathrm{max}}/2)$, which has to be solved numerically for $x_{\mathrm{max}}$.
Thus the maximal intensity scales as
\begin{equation}
    I_{\mathrm{max}}\sim0.4392\Gamma\frac{N^2}{\ln^2 N}.
\end{equation}

\subsubsection{Lossy regime}
In the regime $g_{\mathrm{r}}>1$, particle loss is dominant.
The intensity scales as
\begin{equation}
    I\sim\Gamma x^2N^{1+(1-x)/g_{\mathrm{r}}}
\end{equation}
with $x\coloneqq e^{-\bar t}$.
It becomes maximal at $x_{\mathrm{max}}=2g_{\mathrm{r}}/\ln N$ with
\begin{equation}\label{eq:mf_loss}
    I_{\mathrm{max}}\sim\Gamma\frac{4g_{\mathrm{r}}^2}{\ee^2}\frac{N^{1+1/g_{\mathrm{r}}}}{\ln^2N}.
\end{equation}

\section{Asymptotics beyond the mean-field}
\label{app:stochastic_asymptotics}

The problem admits a matching ansatz combining the large-$N$ results just derived with the treatment in Ref.~\cite{malz2022}. There, the initial superradiant emission at early times is treated as a linearized random process which is then connected to the nonlinear mean-field solution. This method can be directly applied here since only the superradiant fluctuations contribute at leading order in the large-$N$ regime. Fluctuations due to particle loss are subleading. The early superradiant dynamics amplify the initially small ground-state population, while particle loss depletes a large reservoir of emitters.

Thus, we first consider $r=0$ and summarize the result from Ref.~\cite{malz2022}. The superradiant emission induces transitions
\begin{equation}
(N,k)\to(N,k+1), \quad\text{with rates}\quad \Gamma h_{N-k}^{(N)},
\end{equation}
where $k=N-m$ is the ground-state occupation. We choose a matching scale $k_\star$ such that $1\ll k_\star\ll N$ and consider the time evolution until $k$ reaches $k_\star$. In this regime, the dynamics reduce approximately to the birth process
\begin{equation}
k\to k+1,\quad\text{with rate}\quad\Gamma(N-k)(k+1)\simeq\Gamma N(k+1).
\end{equation}

Since $k+1$ appears in the rate instead of $k$, it is easier to think about this process in terms of $X=k+1$. This chain then corresponds simply to a linear pure-birth process and is solved by a geometric distribution
\begin{equation}
    P(X=n,t)=p(1-p)^{n-1},\qquad p=e^{-\Gamma Nt}.
\end{equation}

Note that $P(X=n,t)=p^{(N)}_{N-n+1}(t)$ in our usual notation. It simplifies calculations to now divide out the mean, defining
\begin{equation}
    W_k(t)=\frac{X(t)}{\expval{X(t)}}=X(t)\me^{-\Gamma Nt}=[k(t)+1]\me^{-\Gamma Nt}
\end{equation}

We now find the probability distribution function for $W_k(t)$ by first determining its survival function $P(W_k(t)>w)$. The condition $W_k(t)>w$ is equivalent to asking that $k\geq \lfloor w/p\rfloor$ in the original variables. We can now directly evaluate the survival function
\begin{equation}
    \begin{aligned}
        P(W_k(t)>w)
        &=\sum_{k=\lfloor w/p\rfloor}^\infty p(1-p)^k
        =p\frac{(1-p)^{\lfloor w/p\rfloor}}{1-(1-p)}=(1-p)^{\lfloor w/p\rfloor}.
\end{aligned}
\end{equation}

Since $k\gg 1$ we can also infer that $\Gamma Nt\gg 1$. As such, we may take the limit $\lim\limits_{p\to 0}$ for this equation and obtain
\begin{equation}
    P(W_k(t)>w)\approx \lim\limits_{p\to 0}(1-p)^{\lfloor w/p\rfloor}=\me^{-w}.
\end{equation}

Writing $W$ for the limiting random amplitude, the negative derivative of its survival function gives the probability density
\begin{equation}
f_W(w)=e^{-w},\qquad w\geq0.
\end{equation}

Thus, the waiting-time distribution of the first $k_\star$ photon emissions produces a random amplitude of the exponentially growing ground-state population,
\begin{equation}
    k(t)=W_k(t)\me^{\Gamma Nt}-1\approx W_k(t)\me^{\Gamma Nt},
\end{equation}

Emitter loss, as argued before, can be taken into account completely deterministically, via
\begin{equation}
k(t)=W\,k_{\mathrm{MF}}(t)=W\me^{-rt} \exp\left[\frac{\Gamma N_0}{r}(1-\me^{-rt})\right],
\end{equation}
for $1\ll k(t)\ll N$. Hence we find for the excited-state fraction, including loss
\begin{equation}
  p_W(t)\simeq1-\frac{W}{N_0}\exp\left[\frac{\Gamma N_0}{r}(1-e^{-rt})\right].  
\end{equation}

We now compare this to the full nonlinear mean-field solution $p_{\mathrm{MF}}(t)=(1+\frac{1}{N_0}\exp\left[\frac{\Gamma N_0}{r}(1-e^{-rt})\right])^{-1}$ and find that the two solutions match if we propose 
\begin{equation}
p_{W}(t)=\frac{1}{1+\frac{W}{N_0}\exp\left[\frac{\Gamma N_0}{r}(1-e^{-rt})\right]}
\end{equation}

We may thus obtain the stochastic corrections to the intensity for the previous section from
\begin{equation}
\overline I(\bar t)=\Gamma N_0^2\me^{-2\bar t}\int_0^\infty \me^{-W}p_{W}(\bar t)(1-p_{W}(\bar t))\dd W,
\end{equation}
which may also be expressed, to illustrate the maximum in the three asymptotic regimes,
\begin{equation}
    \overline I(t)=\Gamma N_0^2e^{-2rt}F(z(t)),
\end{equation}
where $F(z)=\int_0^\infty e^{-W}\frac{Wz}{(1+Wz)^2}\dd W$ and $z(t)=\frac{1}{N_0}\exp\left[\frac{\Gamma N_0}{r}(1-e^{-rt})\right]$.

\paragraph{Superradiant regime $g_{\mathrm{r}}<1$: } Once again, for $g_{\mathrm{r}}<1$, the maximum occurs for some finite value of $z(\bar t)$ which does not scale with $N$. One finds that $\overline I(z)=\Gamma N_0^2\left(1-g_{\mathrm{r}}-\frac{g_{\mathrm{r}}\ln z}{\ln N_0}\right)^2F(z)$, and the term proportional to $\ln^{-1} N_0$ can be dropped such that
\begin{equation}
    I_{\mathrm{max}}=\Gamma N_0^2\left(1-g_{\mathrm{r}}\right)^2\max_{z>0}F(z)\approx 0.1957\Gamma N_0^2(1-g_{\mathrm{r}})^2
\end{equation}
where the prefactor is found numerically by maximizing $F(z)$.

\paragraph{Critical regime $g_{\mathrm{r}}=1$: } Once again substituting $x(\bar t)=\ee^{-\bar t}\ln N_0$ we thus find the maximum as
\begin{equation}
    I_{\mathrm{max}}=\frac{\Gamma N_0^2}{\ln^2N_0}
\max_{x>0}\left[x^2F(e^{-x})\right]\approx0.3891\frac{\Gamma N_0^2}{\ln^2N_0}
\end{equation}
where the prefactor is once again found numerically.

\paragraph{Loss-dominated $g_{\mathrm{r}}>1$: } In the loss-dominated regime, superradiant emission never becomes sufficiently amplified. Here $z\to 0$ and $F(z)=z+\mathcal O(z^2)$ corresponding to the deterministic mean field at leading order, see Eq.~\eqref{eq:mf_loss}.

\end{document}